# First Resolution of a Main Sequence G-Star's Astrosphere Using Chandra

**C.M. Lisse[1], S.J. Wolk[2], B. Snios[2], R.L. McNutt, Jr.[1], J.D. Slavin[2], R.A. Osten[3], D.C Hines[3], J.H. Debes[3], D. Koutroumpa[4], V. Kharchenko[2], J.L. Linsky[5], P. Brandt[1], M. Horanyi[5], H.M. Günther [6], E.F. Guinan[7], S. Redfield[8], P.C. Frisch[9], K. Dennerl[10], V. Kashyap[2], K.G. Kislyakova[11], Y.R. Fernandez[12], E. Provornikova[1], M.A. MacGregor[1], C.H. Chen[3], L. Paxton[1], K. Dialynas[13], L. Gu[14]**

---

[1]Space Exploration Sector, Johns Hopkins University Applied Physics Laboratory, 11100 Johns Hopkins Rd, Laurel, MD 20723 carey.lisse@jhuapl.edu, Meredith.MacGregor@pha.jhu.edu, Elena.Provornikova@jhuapl.edu, ralph.mcnutt@jhuapl.edu, pontus.brandt@jhuapl.edu, larry.paxton@jhuapl.edu

[2]Harvard-Smithsonian Center for Astrophysics, 60 Garden Street, Cambridge, MA, 02138 swolk@cfa.harvard.edu, vkharchenko@cfa.harvard.edu, bradfordsnios@gmail.com, jslavin@cfa.harvard.edu, vkashyap@cfa.harvard.edu

[3]Space Telescope Science Institute, 3700 San Martin Dr. Baltimore, MD 21218 osten@stsci.edu, john.debes@gmail.com, hines@stsci.edu, cchen@stsci.edu

[4]LATMOS/IPSL, UVSQ Université Paris-Saclay, Sorbonne Université, CNRS, 11 Boulevard D'Alembert, 78280 Guyancourt, France Dimitra.Koutroumpa@latmos.ipsl.fr

[5]University of Colorado, Boulder, CO jlinsky@jila.colorado.edu, mihaly.horanyi@lasp.colorado.edu

[6]Massachusetts Institute of Technology, Kavli Institute for Astrophysics and Space Research, 77 Massachusetts Avenue, NE83-569, Cambridge, MA 02139 hgunther@mit.edu

[7]Villanova University, Dept. of Astrophysics and Planetary Sci, 800 Lancaster Avenue, Villanova, PA 19085 edward.guinan@villanova.edu

[8] Wesleyan University, Astronomy Department and Van Vleck Observatory, 96 Foss Hill Drive, Middletown, CT 06459, USA sredfield@wesleyan.edu

[9]University of Chicago, Department of Astronomy and Astrophysics, 5640 S. Ellis Ave, Chicago, IL 60637 (deceased)

[10]Max-Planck-Institut für_extraterrestrische Physik, Postfach 1312, Giessenbachstraße, D-85741 Garching, Germany kod@mpe.mpg.de

[11]Department of Astrophysics, University of Vienna, Tuerkenschanzstrasse 17, Wien, Austria A-1180 kristina.kislyakova@univie.ac.at

[12]Department of Physics, University of Central Florida, Orlando, FL 32816 yan@ucf.edu

[13] Center for Space Research and Technology, Academy of Athens, Athens, 4, Soranou Efesiou str., 11527, Papagos, Athens, Greece kdialynas@phys.uoa.gr

[14]SRON Space Research Organisation Netherlands, Niels Bohrweg 4, 2333 CA Leiden, The Netherlands, l.gu@sron.nl

---

Proposed Running Title: **First Resolution of a Main Sequence G-Star's Astrosphere Using Chandra"**

Please address all future correspondence, reviews, proofs, etc. to:

Dr. Carey M. Lisse

Planetary Exploration Group, Space Exploration Sector

Johns Hopkins University, Applied Physics Laboratory

SES/SRE, Bldg. 200, E206

11100 Johns Hopkins Rd

Laurel, MD 20723

240-228-0535 (office) / 240-228-8939 (fax)

Carey.Lisse@jhuapl.edu

## Abstract

We report resolution of a halo of X-ray line emission surrounding the Zero Age Main Sequence (ZAMS) G8.5V star HD 61005 by Chandra/ACIS-S. Located only ~36 pc distant, HD 61005 is young ($100^{+50/-50}$ Myr), x-ray bright (~300x Solar), observed with nearly edge-on geometry, and surrounded by Local Interstellar Medium (LISM) material denser than in the Sun's environs. HD 61005 is known to harbor large amounts of circumstellar dust in a dense ecliptic plane full of mm-sized particles plus attached, extended "wing like structures" full of micron sized particles, which are evidence for a strong LISM-dust disk interaction. These properties aided our ability to resolve the system's ~220 au wide astrosphere, the first ever observed for a main sequence G-star. The observed x-ray emission morphology is roughly spherical, as expected for an astrospheric structure dominated by the host star. The Chandra spectrum of HD 61005 is a combination of a hard stellar coronal emission (T~8 MK) at $L_x$~6 x$10^{29}$ erg/sec, plus an extended halo contribution at $L_x$~1x$10^{29}$ erg/sec dominated by charge exchange (CXE) lines, such as those of OVIII and NeIX. The Chandra CXE x-ray morphology does not track the planar dust morphology but does extend out roughly to where the base of the dust wings begins. We present a toy model of x-ray emission produced by stellar wind (SW)-LISM CXE interactions, similar to the state of the young Sun when it was ~$10^8$ yrs old (Guinan & Engle 2007), and transiting through an ~$10^3$ times denser part of the interstellar medium (ISM) such as a Giant Molecular Cloud (Stern 2003, Opher & Loeb 2024).

**1. Introduction.** Stars shine in the x-ray due to photon emission from the hot (~MK) collisional plasmas in their surrounding coronal atmospheres (Testa 2010). However, stars also produce a low level of x-ray emission over a large volume, as the ionized, high pressure stellar winds flowing out from their coronae blow a bubble/cavity (termed an "astrosphere") in the local galactic ISM (Parker 1965; Dialynas *et al.* 2017; Opher *et al.* 2015, 2020, 2021), and charge exchange processes that transfer energy and pressure between inflowing ISM neutral gas and the stellar wind ions create x-rays (Cravens 2000, *et al.* 2001; Koutroumpa *et al.* 2009), especially near the excavated bubble's outer reaches (termed an "astropause"; Linsky *et al.* 2019).

Charge exchange (or charge exchange collision) is a plasma process in which an ion of charge state $+n$ ($A^{+n}$) collides with a neutral atom or molecule ($B$), resulting in the neutral atom losing and the ion gaining one or more electrons, i.e., $A^{+n} + B \xrightarrow{yields} A^{+(n-1)} + B^{+}$. The cross section for this process is very large, 10 – 100 times larger than the hard sphere collision radius for the two particles. Charge exchange processes in astrophysics are thought to be common, occurring wherever a hot ionized plasma encounters a cold neutral gas, as in solar wind interactions with cometary or planetary atmospheres (Lisse *et al.* 1996, 2004; Dennerl *et al.* 2012), the solar wind plowing into the very local interstellar medium (VLISM) at the heliopause (Wood *et al.* 2005, 2021), in novae outbursts (Mitrani *et al.* 2024, 2025), in hot supernova blast waves shocking the galactic ISM, or in starburst galaxies and galaxy clusters (Gu & Shah 2023). In our solar system's astrosphere (aka the heliosphere), the process is easily detected by the characteristic x-rays emitted by the ion which receives the electron(s), as these are typically captured into excited n=4 to 6 principal quantum number states, which then relax to their ground state via energetic photon emission (Cravens 1997, Kharchenko *et al.* 2003, Koutroumpa *et al.* 2009).

Searches for exosystem astrospheric charge exchange x-ray emission have been conducted since the discovery of solar-wind charge exchange (SWCX) from comets, the planets, and the heliosphere in the 1990s and early 2000s (Dennerl *et al.* 2012 and references therein), but none have been successful until recently (Kislyakova *et al.* 2024), despite decades of Lyman-α measurements of charge-exchange-produced neutral hydrogen "walls" (Wood *et al.* 2005, 2021; Linsky *et al.* 2019). According to current heliospheric models (e.g., Dialynas *et al.* 2017, Pogorelov *et al.* 2017,

Kornbleuth *et al.* 2021), charge exchange processes **should** occur in the outer boundaries of all stellar systems as stellar winds blow bubbles in the galactic ISM, but x-ray production is at too low a rate to be detected parsecs away versus the cosmic background and the host star's coronal emission using current instrumentation (Wargelin & Drake 2001).

HD 61005 (G9 V; Fig. 1) is known as both a ROSAT All-Sky Survey (RASS) X-ray and ultraviolet (XUV) source and a highly structured and dynamic young Sun-like debris disk system. Also called "The Moth" due to its unique morphology (Hines *et al.* 2007), HD 61005 is a $T_{eff}$ = 5500 ± 50 K, $L_\star$ = 0.583±0.048 $L_{Sun}$, $M_\star$ ~0.90 $M_{Sun}$, $R_\star$ =0.840±0.038 $R_{Sun}$ , d = 36.4 pc late G-star (Hines *et al.* 2007, Desidera *et al.* 2011, Brown *et al.* 2018) located at the edge of the Local Interstellar Cloud (LIC; Redfield & Linsky 2000) in the rough direction of Epsilon Canis Majorus (the strongest source of local XUV emission dominating the Local Bubble (LB); Vallerga *et al.* 1993, Vallerga & Welsh 1995, Gry & Jenkins 2001). 2MASS photometry (Desireda *et al.* 2011) demonstrates a normal main sequence color and luminosity when compared to typical values for a Gyrs-old main sequence G9V star scaled to its distance[1]. This also implies that the optical extinction is not large from the Earth towards HD 61005, although it may be located within the nearby Blue, G, or Cet Clouds with $n_{neutral}$ (mostly H-atom) densities upwards of 100 cm$^{-3}$ (Redfield *et al.* 2008, Peek *et al.* 2011, Meyer *et al.* 2012, Section 4; compare this to the ~0.1 cm$^{-3}$ of the LISM near the Sun). Age estimates for the system, including its stellar rotation period $P_{rot}$ = 5.04±0.04 days (Desidera *et al.* 2011) and association in the Argus stellar association, are in the 100 ± 50 Myr range, and all agree that it is close to or at ZAMS (Desireda *et al.* 2011, Casagrande *et al.* 2011). No planet detections have been reported for the system to date, there is no evidence for a nearby binary stellar companion, nor is there any detected misalignment between the star's rotation axis and the disk (Desireda *et al.* 2011).

Sourced by material emitted from rocky and icy planetesimals, circumstellar debris disks bear the imprints of endogenous planetary formation and evolution processes (Zuckerman & Song 2004, Chen *et al.* 2006, Hughes *et al.* 2018). A few disks have additional bowshock-like structures, as if they were plowing through a heavy external interstellar wind (e.g. Delta Velorum, Gaspar *et al.* 2008). One of the most pronounced of these systems, HD 61005, shows a pronounced set of

[1] (https://www.pas.rochester.edu/ ~emamajek/EEM_ dwarf_UBVIJHK_colors_Teff.txt)

"wings" composed of micron sized grains, as first revealed by HST/NICMOS near-infrared (NIR) observations (Hines *et al.* 2007). Using knowledge of the primary star's relative space motion,

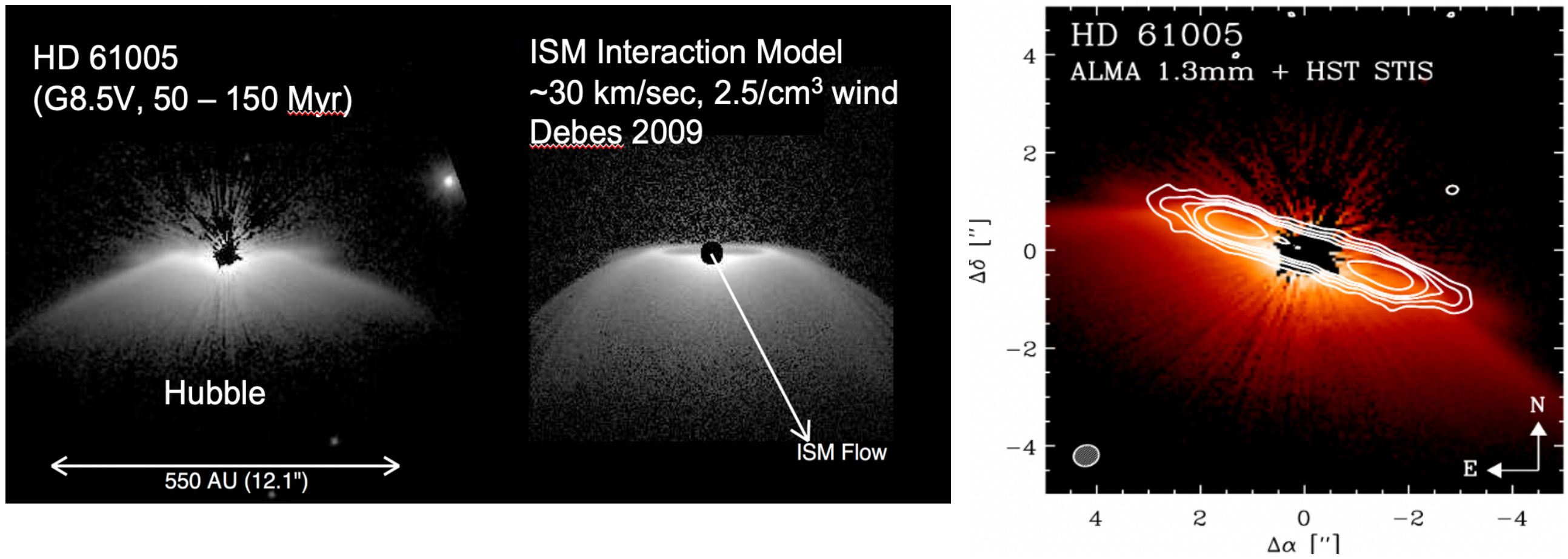

**Figure 1 - Archival HST/STIS near-infrared imagery of HD 61005. (Left)** The swept-back wings of the outer disk can be clearly seen in contrast to the bright central flat disk running left-right in the center of the image (Schneider *et al.* 2014). **(Center) Debes *et al.* 2009 model of the system's dust structure** produced by invoking ISM wind ram pressure perturbations of circumstellar dust orbits. Using the image and space motion analysis determined in Hines *et al.* 2007, the system is modeled to be observed perfectly edge-on, with the ISM flow vector in the plane of the figure. **(Right).** ALMA sub-mm imagery of the large (mm-cm) sized dust particles in the system (white contours, MacGregor *et al.* (2018)) overlaid on top of HST/STIS data (red imagery, Schneider *et al.* 2014).

Debes *et al.* (2009) and Maness *et al.* (2009) successfully reproduced these wing-like structures using models incorporating the ram pressure effects of a locally dense interstellar medium (ISM; ($n_{ISM,\ HD61005} \sim 100\ cm^{-3}$ vs. $\langle n_{ISM,\ Average} \rangle \sim 0.1\ cm^{-3}$) on dusty material orbiting the star, for a system with an ecliptic plane presenting itself face-on to the in-streaming ISM (Figure 1a)[2]. Effects due to a dense Very Local ISM (VLISM) are consistent with the system's designation as a G9Vk (k for unusually strong ISM absorption lines) in the literature (Gray *et al.* 2006). VLT mid-infrared (MIR) observations (Olofsson *et al.* 2016) and ALMA sub-mm observations (MacGregor *et al.* 2018) have recently added to this picture the presence of a dense edge-on disk containing very large amounts (0.05 – 0.40 $M_{Earth}$) of mm- to cm-sized dust grains.

If inflowing ISM winds are so demonstrably responsible for shaping the HD 61005 dust disk, they could also create signature XUV photons from charge exchange interactions in the HD 61005 astrosphere and astropause boundaries (and perhaps even in a "downstream Mira-like tail", Wareing

[2] For completeness, we note that other groups (e.g., Buenzli *et al.* 2010, Esposito *et al.* 2016, Olofsson *et al.* 2016, Lin and Chiang 2019) have suggested instead that HD 61005's wings are a solely endogenous phenomenon, due to the influence of inner planets on highly eccentric perturbing orbits interacting with massive planetesimal collisions. The results of the Chandra observations reported here strongly support the dust disk-ISM interaction model.

2008) coincident with the bowed-back regions of the disk. Because of the high density of the neutral interplanetary and local ISM material surrounding the HD 61005 host-star, CXE rates could be much higher than those known to be occurring in our own heliosphere and heliosheath/heliopause solar wind-ISM interaction regions (e.g., Koutroumpa *et al.* 2009; Gurnett *et al.* 2015; Dialynas *et al.* 2017, 2019), given enough stellar wind flux. Fortunately, young ZAMS stars typically have stellar winds and XUV emission hundreds of times higher than their mature mid-main sequence counterparts (Ribas *et al.* 2005, Wolk *et al.* 2005, Guinan & Engle 2007, Tu *et al.* 2015).[3]

HD 61005 with its bright circumstellar dust structures thus presented an interesting opportunity for high spatial resolution x-ray study with Chandra, since here was a 50-150 Myr old system with reports that it had been detected by ROSAT and GALEX at the $L_x = 10^{30}$ erg sec$^{-1}$ level (Desireda 2011; Fig 1b). With a reported possible 12" spatial extent of the system, HD 61005 was well located for Chandra imaging, being close enough to Earth to appear x-ray bright, but far enough away that it could be resolved with an estimated extent covering 20 to 30 Chandra **Advanced CCD Imaging Spectrometer (**ACIS)-S pixels. A very short 4 ksec Chandra/HRI 7' off-axis exposure in 2014 did detect it as an $L_x \sim 3 \pm 2 \times 10^{30}$ erg sec$^{-1}$ source at low confidence, roughly consistent with the luminosity expected for an ~100 Myr old G8.5V object (Osten & Wolk 2015).

With all these considerations in mind, time was requested and received to re-observe HD 61005 in Chandra Cycle 21 using ACIS-S, supposing that extended x-ray emission structures and spectroscopy due to dust-ISM or stellar wind-ISM interactions would be found. This paper reports the results of these observations.

## 2. Observations.

In this Section we present the circumstances and results of our 2021 Chandra observations of the HD 61005 system.

---

[3] Our Chandra observations indeed find the system to be x-ray bright ($L_x \sim 10^{30}$ erg sec$^{-1}$ for HD 61005, ~300 times $L_{x,\,Sun}$). This high x-ray luminosity and stellar wind flux also has important consequences for the circumstellar dust in the system - the mm-sized dust detected by ALMA in HD61005 (Fig. 1a) should be removed on 0.1 Myr timescales via stellar wind pressure (very quickly versus the usual 1-10 Myr timescales for radiative P-R drag mechanisms clearing debris disks; Chen *et al.* 2005, Lisse *et al.* 2017). This implies that there must be strong, active sourcing of HD 61005's dust disk to counterbalance the stellar wind sink. Sputtering of circumstellar material (like young KBOs) near the astroshock by an ~30 km sec$^{-1}$ ISM wind (Debes 2009) could produce significant amounts of fresh dust, including the micron-sized particulates of the dramatically swept back disk wings. We revisit this point in Sections 4.2, "Direct Implications", and Section 7, "Open Questions", below.

**2.1 X-ray Photometry & Luminosity.** Chandra ACIS-S imaging spectroscopy measurements of the HD 61005 system were taken in 2 visits of 33 and 34 ksec exposures. The first observations started at 23-Feb-2021 UT (Chandra ObsID 22348) and the second set on 25-Feb-2021 UT (Chandra ObsID 22349). For the visits, no gratings were used, and the star was centered as close as possible to on-axis of the S3 chip to minimize image PSF distortions. We re-processed the Chandra data with CIAO version 4.14 (Fruscione *et al.* 2006), which applied the energy dependent sub-pixel event repositioning. The total on-target observing time was 67 ksec, in which ~2198 total raw photons over the 0.45 - 2.0 keV band were detected using a source extraction radius of 6″. The background in each visit was estimated from the entire S3 ¼ subarray's worth of off-source signal: 2.57 total [2.57/(256*1024) = $9.8 \times 10^{-6}$ cps $pixel^{-1}$] and 2.78 cps total [2.78/(256*1024) = $1.1 \times 10^{-5}$ cps $pixel^{-1}$]. Net total count rates for HD 61005, the brightest object in the field, of 0.031 and 0.032 cps were observed, so pile-up was negligible for the measurements. After subtracting the small background contribution, we determine a total of 2195 source counts from both pointings.

Examination of the photometry time series (light curve) of the Chandra observations of HD 61005 does not show any obvious trends above the noise, which is not surprising since we have only covered 16% of the reported $P_{rot}$ = 5.04 days (Desireda *et al.* 2011) rotational period of the star, and the probability of observing a stellar flare for a star, even a ZAMS late-type star, is small in only 67 ksec of exposure time.

**2.2 X-ray Imaging.** On the ACIS-S3 chip, one x-ray-bright source in the raw HD 61005 imagery was detected, near the S3 aimpoint centered on HD61005's celestial position (Fig 2). This source appeared roughly axially symmetric, but significantly extended as opposed to the usual stellar objects observed by ACIS, such as the few other fainter point sources we could find in the 2 visits (Fig 2). Image deconvolution did not find any evidence for multiple point sources at HD 61005's location.

In the 2' x 2' subarray field around the HD 61005, a companion search was performed for potential nearby x-ray sources in the field greater than 3 times the observed image background level of ~$5 \times 10^{-7}$ cps per pixel. [Given an ACIS-S 90% encircled energy radius of 2.5 pixels, this implies a $3\sigma$ upper limit for any x-ray object in the field of $9.8 \times 10^{-6}$ cps (approximately equivalent to $L_x \sim$ 4 x $10^{24}$ erg $s^{-1}$, assuming $kT$ = 0.40 keV) for an object at the 36.4 pc distance of HD 61005.] There is evidence in our data for one other point source with ~45 times lower ACIS count rate at ~1.3

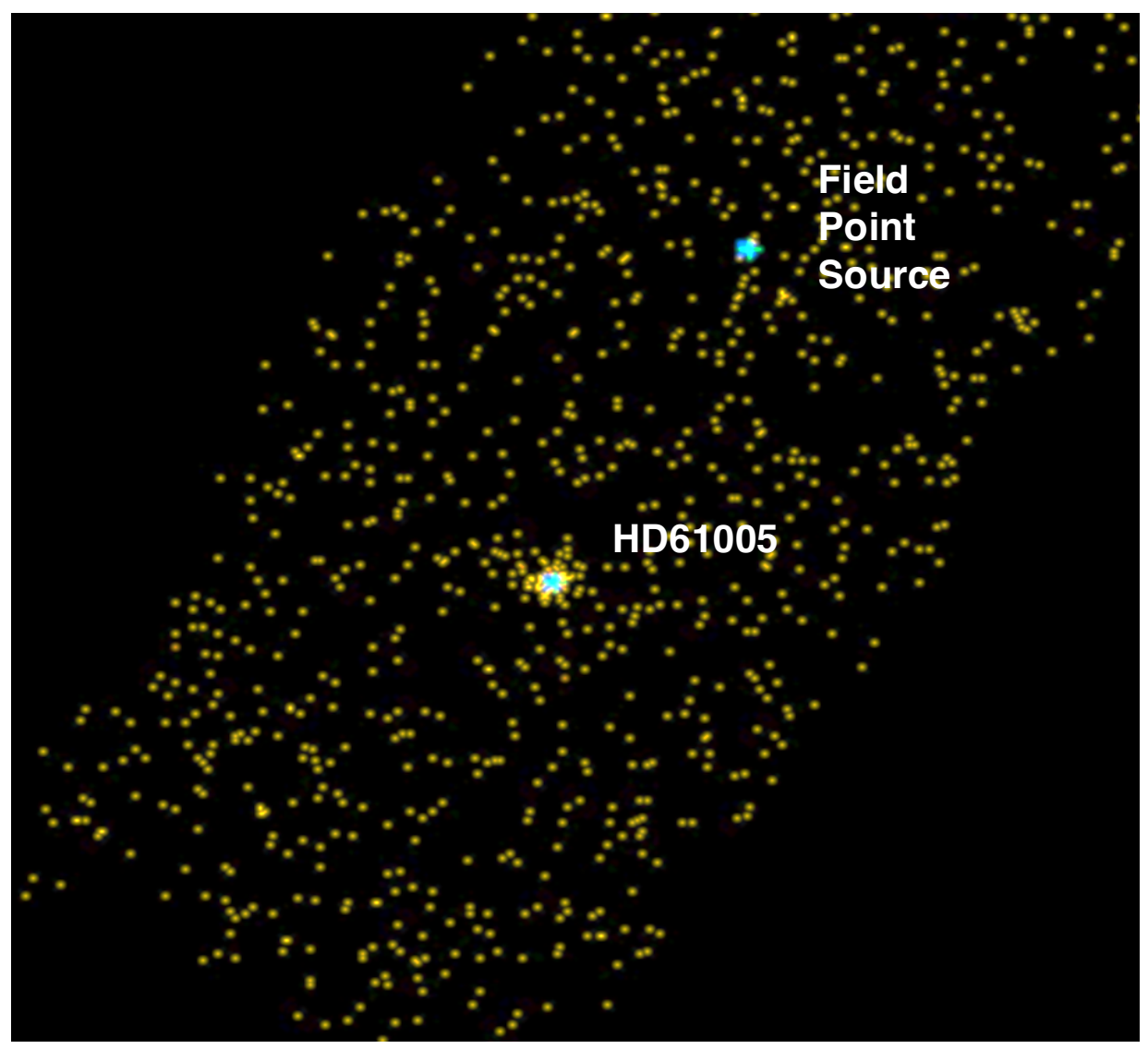

**Figure 2 - Chandra ACIS-S x-ray imagery of HD 61005 from Visit #2 (UT 2021-02-25T03:15:57). (Left)** 256 x 1024 pixel, "¼ subarray" S3 chip image of the HD61005 field, with bright HD 61005 centered at the "sweet spot" of the S3 chip. North is up, and East is to the left in this image. Note also the statistically significant detection of another, much fainter object (~80x fainter) approximately 1.3' N and 0.7' West of HD 61005. The field object is stellar, i.e. it exhibits the same aperture photometry radial dependence as point sources Tau Ceti and Beta Hyi (see Section 3 & Figs. 4-5). By contrast, the HD 61005 source shows the same ~2.5 pixel radius core plus extended emission levels out to at least 6 pixels from its center of brightness.

arcmin separation (07h35m47.34s -32d 12m12s for HD 61005 vs. the new "field source" at 07h35m44.2s -32d10m57s), but there is no current evidence that this is a separate, distant 2nd member of the HD 61005 system – it could just as likely be a low stellar mass member (a late M-star or brown dwarf) of the young Argus association (Desidera *et al.* 2011) – or, if associated with GAIA source ID 5592237948446238848 reported within 1.2" of HD 61005 (Brown *et al.* 2018), then it has significant parallax and proper motion at ~1900 pc and $T_{eff}$ ~5000 K, log g=3.5, making it an unrelated, distant early K-giant. This result is consistent with the marginal detection of a few small, faint RASS sources in the 4.2′ x 4.2′ field centered on HD 61005 (Voges *et al.* 1999; Fig 1b).

**2.3 X-ray Spectroscopy.** From a total of 2198 detected events, a total HD 61005 spectrum was extracted over both visits within an r = 6 pixel region (of total radius 6.25", after including the width of the central r = 0 half-pixel) radius circular aperture. The total combined spectrum, averaged into 14 eV bins, is shown in Figure 3a, and its best-fit 1-temperature APEC model spectrum in Fig. 3b. Assuming solar metallicity and a G9V's photospheric temperature and log g ($T_{eff}$ = 5408 - 5503 K, log g ~ [4.35 - 4.59], Fe/H ~ [-0.13 to 0.08]; Lepine *et al.* 2003, Casagrande *et al.* 2011, Maldonado *et al.* 2012, Brewer *et al.* 2016, Luck 2018, Chavero *et al.* 2019, Casali *et al.* 2020), the core spectrum can be fit by an emission model for collisionally-ionized diffuse gas (MEKAL "Mewe-Kaastra-Liedahl" code combined with the Astrophysical Plasma Emission Code (APEC) using the Astrophysical Plasma Emission Database (APED)) (Phillips *et al.* 1999, Smith *et al.* 2001) calculated using APEC (atomic database) ATOMDB v2.0.1 in the on-line X-ray spectral fitting program XSPEC (Arnaud *et al.* 1999) modeling package. The best-fit core model appears to be hard versus the Sun, with an APEC temperature of 0.65 ± 0.12 keV (7.5 ± 1.4 x $10^6$ K). While ~6 times hotter than the effective coronal temperature of the 4.5 Gyr-old Sun (~0.10 keV, or 1.2 MK), it is consistent in temperature with other ~100 Myr solar type stars of similar $L_X$ (Suchkov *et al.* 2003, Scelsi *et al.* 2005, Telleschi *et al.* 2005, Tu *et al.* 2015). The continuum-dominated shape of the spectrum, peaking at ~0.9 keV, is also similar to the examples shown in Telleschi *et al.* 2005 in their X-ray spectral survey of **young** solar-type stars. The high coronal temperature means that species like OVIII, FeXVII, and NeIX/NeX should dominate HD 61005's stellar wind, unlike the OVII + CIV/V + NV/VI + NeIX species that dominate for Sun's wind (Bodewits *et al.* 2007), e.g., we can expect OVIII/OVII to be ~10 for HD61005, not the ~0.1 found in the Sun's wind (Section 3.2).

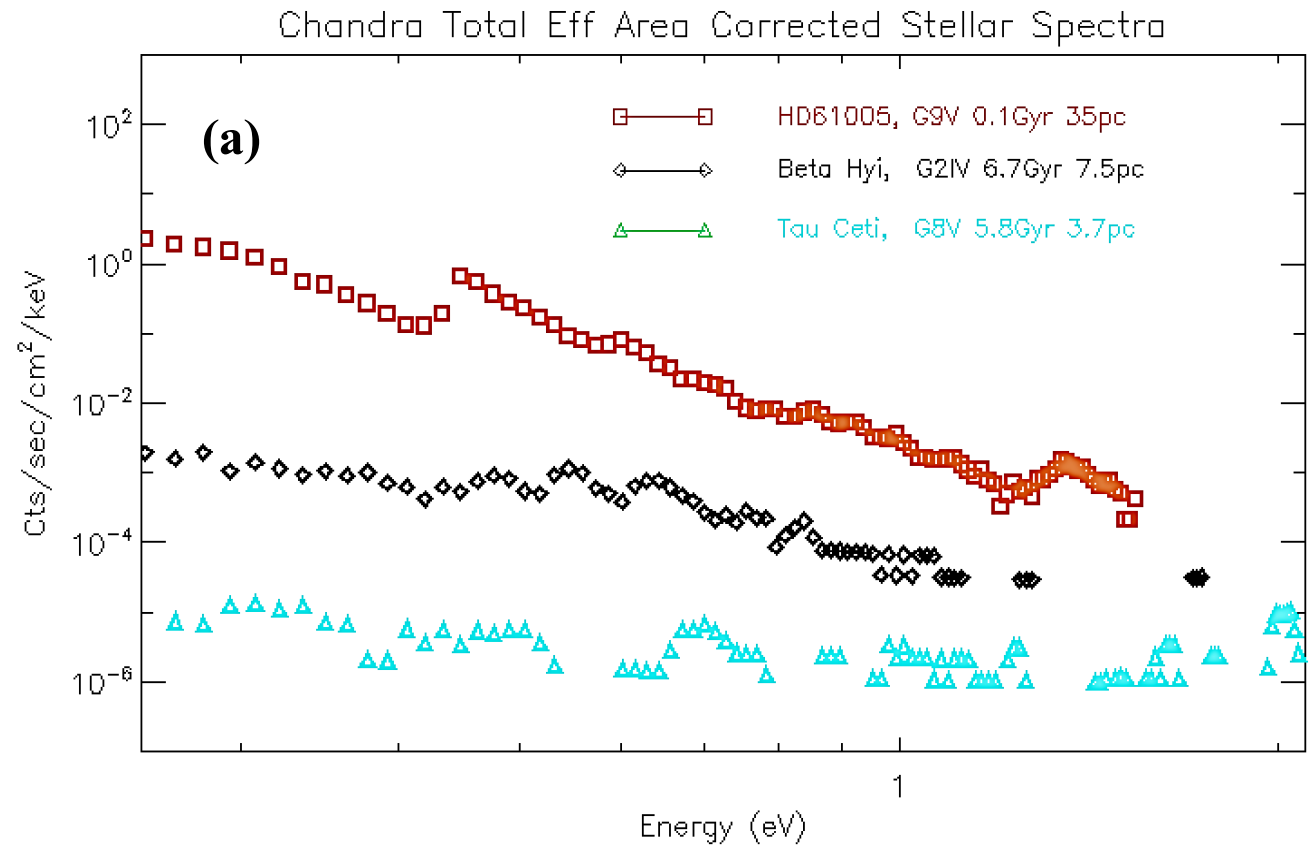

**Figure 3a - Comparative Chandra/ACIS-S spectra for HD 61005, Tau Ceti, and Beta Hyi.** X-ray spectral fluxes are for the 3 sources after background removal, correction for the relative on-target times [$t_{,\mathrm{HD\ 61005}}$, $t_{\mathrm{OTT,Tau\ Ceti}}$, $t_{\mathrm{OTT,Beta\ Hyi}}$ = 67, 67, 51 ksec] of each observation, correction for distance effects assuming $F_x \sim 1/\mathrm{distance}^2$, and correction for the time dependent ACIS-S effective area (as tabulated yearly at https://cxc.harvard.edu/cgi-bin/prop_viewer/build_viewer.cgi?ea). An artifact due to the carbon K-alpha edge at ~284 eV can be clearly seen which affects low energy measurements below 450 eV (see Appendix). Thus all analysis discussed in this paper are from photons of energy 0.45 keV and above, and the x-ray luminosities quoted are for the 0.45 – 2.0 keV range. HD 61005 is $10^3$ – $10^4$ times brighter than the two older stars.

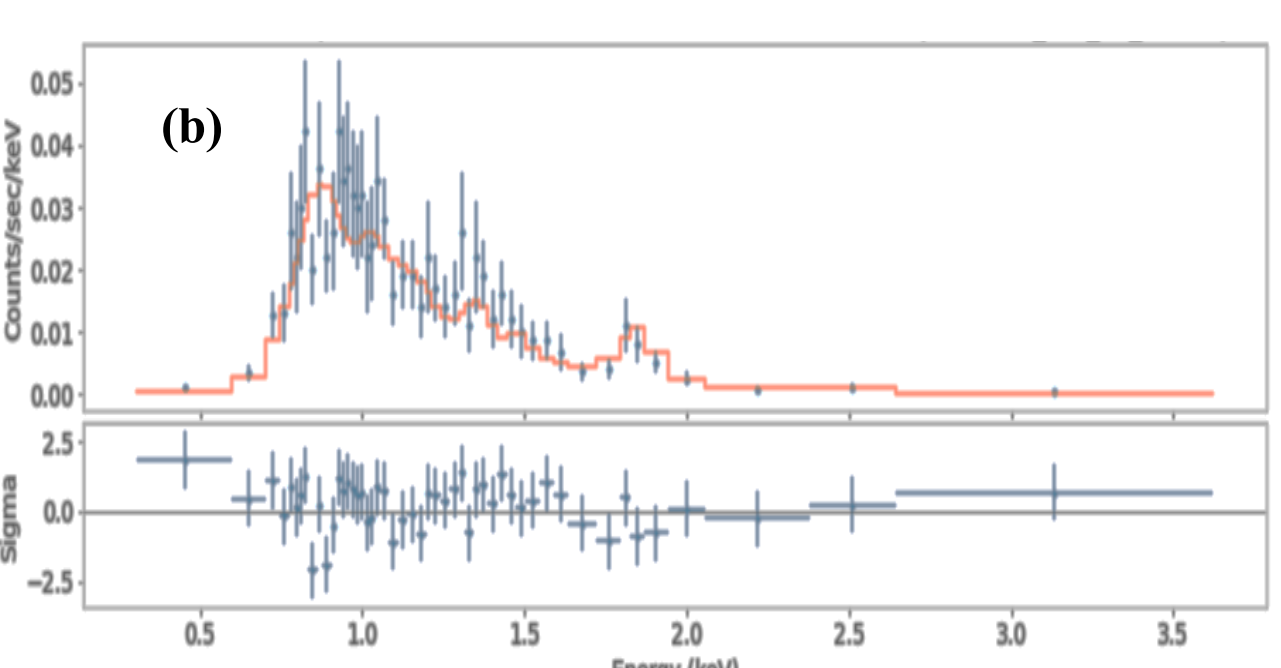

**Figure 3b – Chandra spectral model fit.** Best-fit 1-temperature (0.65 keV/7.5 MK) solar abundance APEC model spectrum assuming hydrogen absorption due to an $N_H = 10^{20}$ cm$^{-2}$ column (orange curve) versus the ACIS-S data with statistical error bars (blue points) for the 90% of events detected within 3 pixels (1.5") of the center of brightness, i.e. for stellar coronal x-ray photons. The inset below the main plot shows the statistical significance of the (Data - Model) residuals.

From the spectrum + model fit, we estimate a total HD 61005 system 0.45 – 2.0 keV X-ray luminosity of $L_x = 0.7 \times 10^{30}$ erg $s^{-1}$ and a mean coronal temperature of 7.5 MK, ~3 to 8 times hotter than the Sun's 0.97 to 2.57 MK (Peres 2000) (Güdel 2004, Aschwanden 2006). The total flux detected is $1.4 \times 10^{-13}$ ergs $cm^{-2}$ $s^{-1}$, from 0.45 to 2.0 keV. The derived hydrogen column of $N_H \sim 1.7 \times 10^{20}$ $cm^{-2}$ from the modeling is an order of magnitude higher than the $\sim 1.4 \times 10^{19}$ $cm^{-2}$ expected for a stellar source ~36 pc distant separated by a typical average intervening interstellar H density of 0.1 particles $cm^{-3}$, suggesting that there is an ISM density enhancement along the line of sight (Section 5).

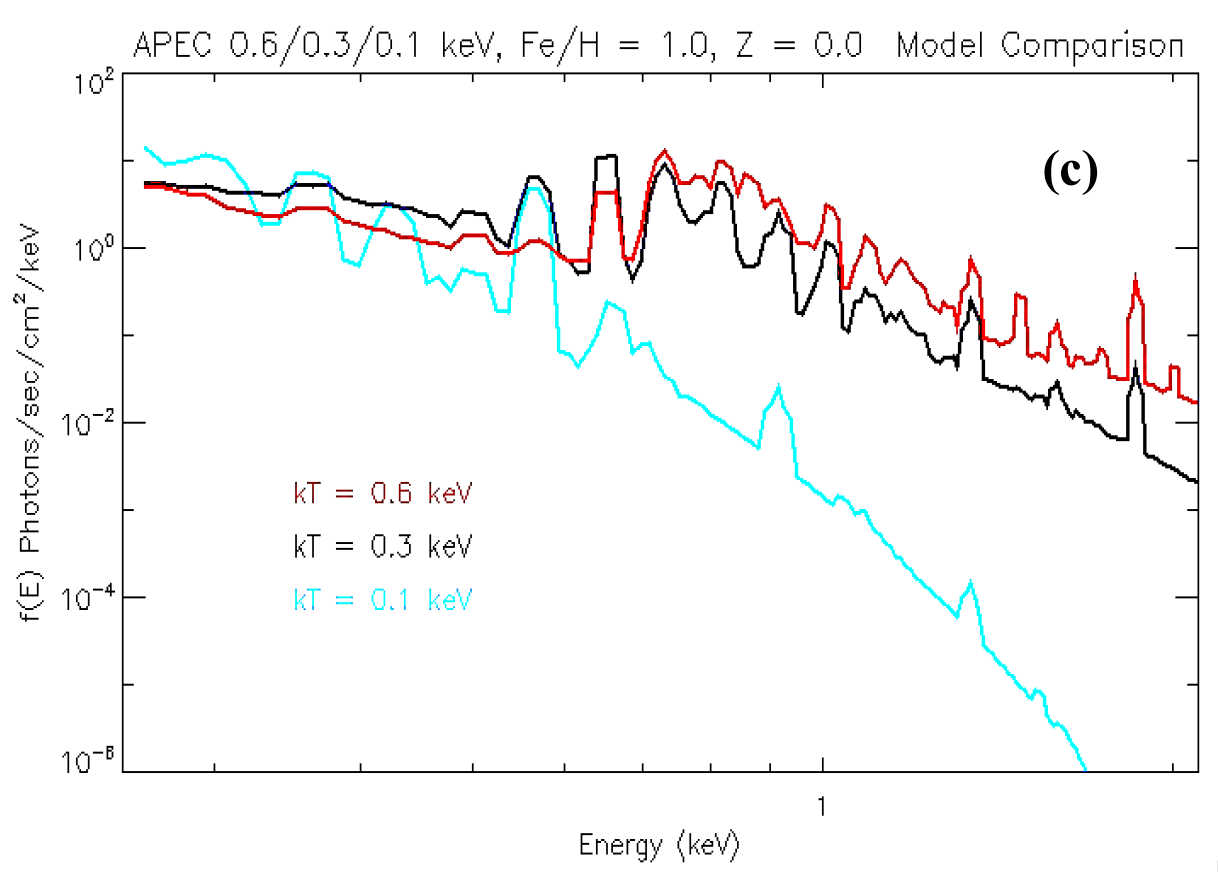

**Figure 3c - Chandra ACIS-S spectroscopic model comparison of hot young stars like HD 61005 vs cooler, mature stars like Beta Hyi and Tau Ceti**. HD 61005, at coronal temperature kT ~ 0.65 keV, is predicted to produce a much brighter and harder spectrum than the mature stars' kT ~ 0.1 keV. This is consistent with HD61005 being a 100 Myr +/- 50 Myr ZAMS star versus the ~6 Gyrs age for Beta Hyi and Tau Ceti (and vs. the ~4.6 Gyrs age and kT~ 0.1 keV found for the Sun). Note the large extent, 8 orders of magnitude, of the logarithmic scale on the vertical axis of each plot.

In Figure 3a we have also plotted the archival measured Chandra/ACIS spectrum for ~4-to-12 Gyr old G8V main sequence star Tau Ceti (Korolik *et al.* 2023) and see that HD 61005 evinces a much brighter, harder spectrum. In Figure 3c, we present the modeled, expected spectra for HD 61005 and Tau Ceti, and see that this is consistent - the hotter and brighter HD 61005 spectrum is as expected (Tu *et al.* 2015, Pastor 2017) for a much younger and more rapidly rotating star like HD 61005 ($P_{rot} \sim 5.04$ days, Age 100 +/- 50 Myrs; Desireda *et al.* 2011, Mamajek, priv. comm.) versus an at least ~100× older star like Tau Ceti ($P_{rot} \sim 46$ days, Korolik *et al.* 2023) or Beta Hyi ($P_{rot} \sim$ 28 days; Ribas *et al.* 2005).[4] Because of its similar age and mass, we expect HD 61005 to act similarly to Güdel *et al.* 1997's and Guinan and Engle 2007's youngest solar twin from their Sun in Time program - the EK Dra system - which is markedly hotter and about $10^3$ times brighter than today's Solar emission activity (Figs 3d and 3e). Our new estimates for $L_x$ versus age are also consistent with the updated predictions of Tu *et al.* (2015; Fig 3f).

[4]Using Solar System X-ray emission as the best proxy for understanding the spectroscopy of exo-solar system emission, said emission should be dominated by solar coronal emission, with minor contributions from charge exchange between highly stripped solar wind heavy ions and neutral gas atoms in atmospheres and the in-streaming ISM wind. As such, the X-ray emission from a stellar system is normally estimated by modeling the emission from a MK hot plasma of the host star's metallicity (ignoring any coronal FIP effects), with mean plasma temperature set by the star's coronal heating as a function of its mass and rotation period/age driving increased flaring and magnetic reconnection (Parker 1988, Testa *et al.* 2010, Raouafi *et al.* 2023). In general, younger stars with higher angular momentum are rotating much rapidly and producing hotter coronae and non-thermal XUV radiation via more vigorous magnetic reconnection (Fig 3d and 3e).

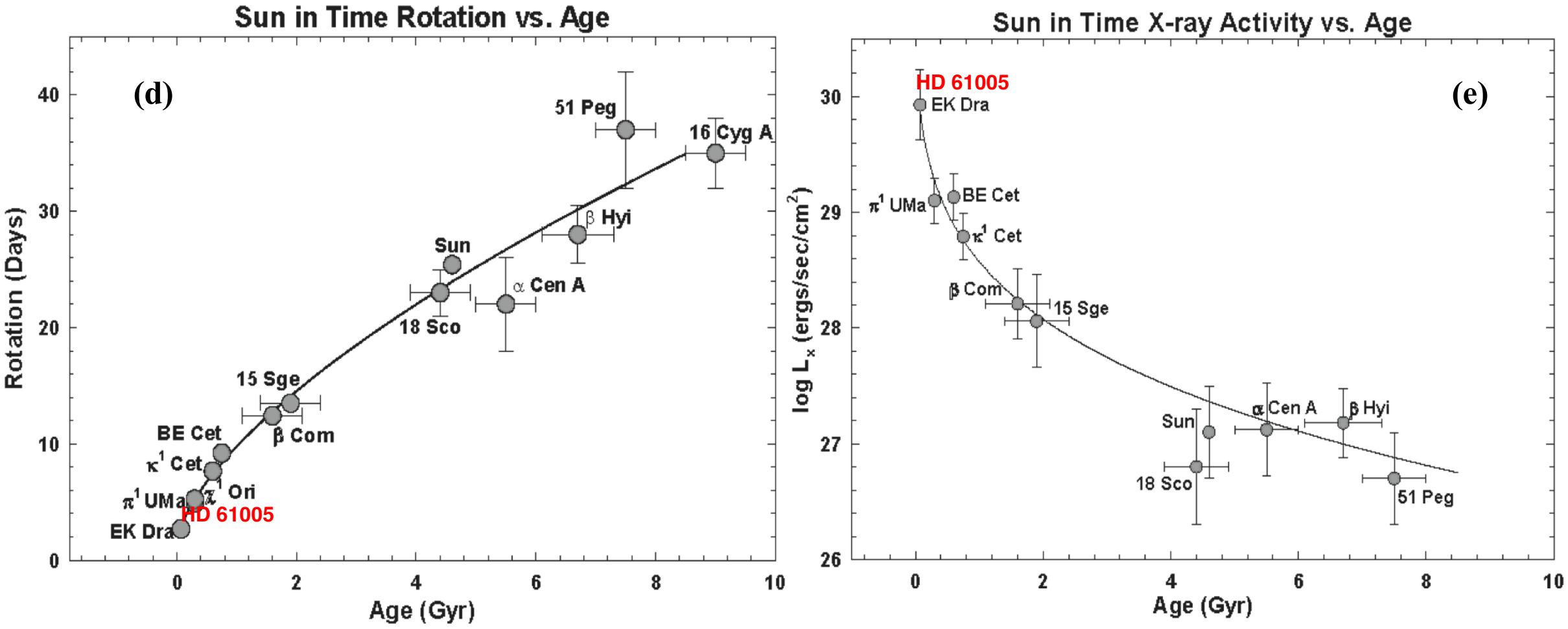

**Figure 3d – Measured rotation rate vs stellar age for the Sun and several close solar analogues.** The solid curve is a simple power law fit modeling $P_{rot} \sim Age^{0.6}$. **(e) – As measured XUV luminosities for EK Dra, π¹ Uma, π¹ Ceti, Beta Com,** and **Beta Hyi, all close solar analogue stars.** Notice the factor of ~$10^3$ higher flux between EK Dra (~ same kind of high emission activity as HD 61005) and β Hyi (~ same kind of low emission state as Tau Ceti). After Guinan & Engle (2007).

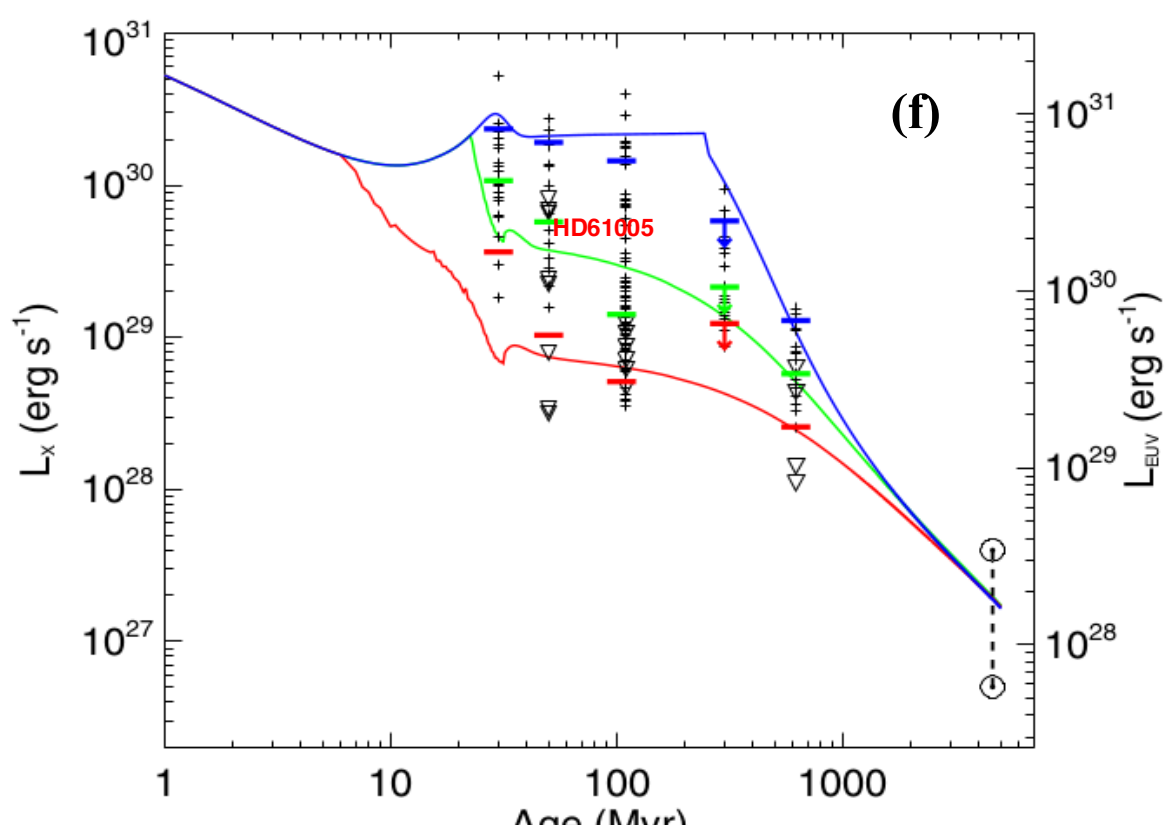

**Figure 3f** – Update of the $L_x$ vs Age plot including an order of magnitude more G-stars and predicted low (red), median (green) and high (blue) x-ray flux models, after Tu *et al.* 2015. HD61005's measured flux lies comfortably between the predicted medium and high cases for an ~100 Myr star.

In the Analysis section below (Section 3), we discuss more details of HD 61005's x-ray line emission and its spatial distribution, both of which differ from the expected result for an unresolved, point source-like coronal x-ray source like EK Dra and Tau Ceti. In Section 4 we discuss the implications of the derived hydrogen column.

**3. Analysis & Models.** Here we present first-order analysis results of our Chandra observations of HD 61005 in comparison to x-ray observations of other stars, derived using algebraic scaling calculations.

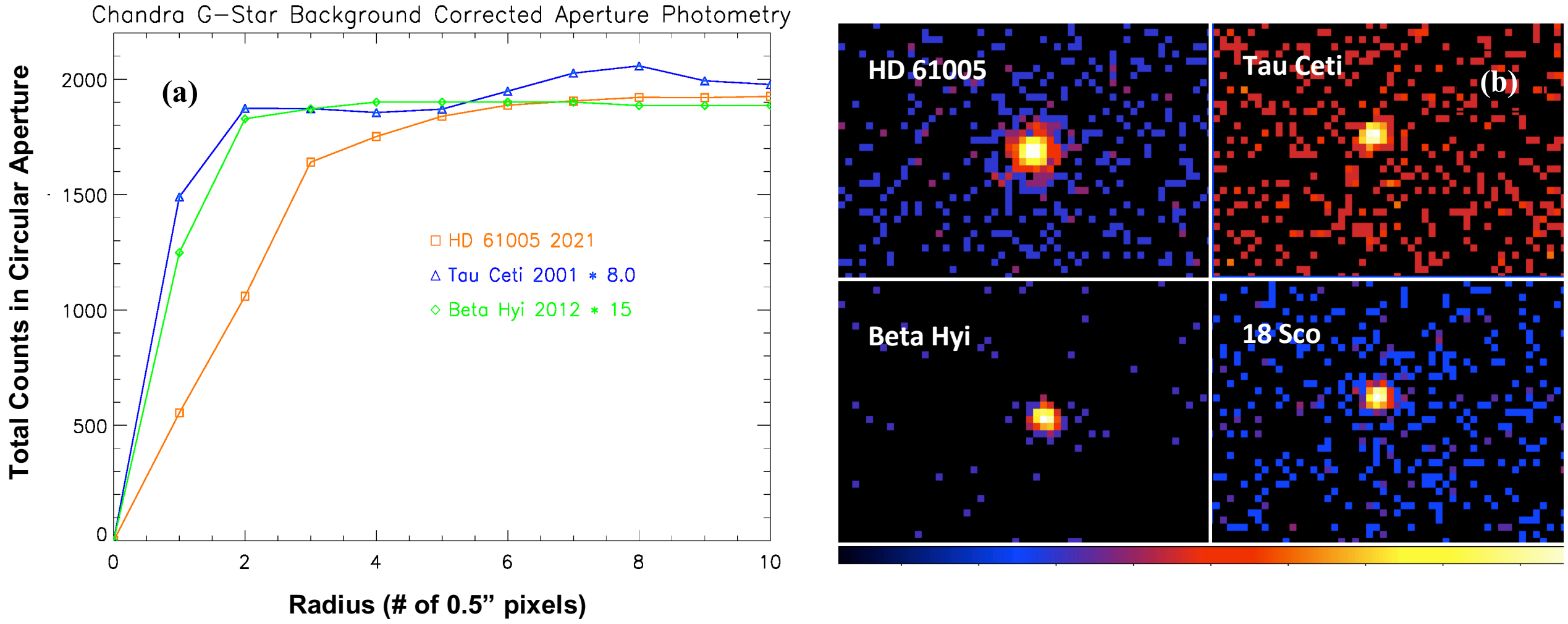

**Figure 4 – Chandra ACIS-S azimuthally averaged radial photometry of HD 61005** (HD 61005, G9V, 40-130 Myr, 36 pc) versus that of two other nearby bright G-star x-ray observed by ACIS-S, Tau Ceti (G8V, 5800 +/- 1200 Myr, 3.7 pc) and Beta Hyi (G2IV, 6700 +/- 1500 Myr, 7.5 pc). **(a)** 0.45 – 2.0 keV aperture photometry for HD 61005, Tau Ceti, and Beta Hyi vs radius. The point source structure found in all 3 systems out to ~1” (2 pixels) is clearly seen. Only HD 61005 shows additional flux above background that is statistically significant out to at least 3.0” (6 pixels). **(b)** Clockwise from upper left: **ACIS-S images of HD 61005, Tau Ceti, Beta Hyi, and 18 Sco**, another solar analogue imaged by Chandra. (**c**) **Replotting of all 0.45 – 2.0 keV photons detected within a 50 x 50 pixel region centered on Beta Hyi, Tau Ceti, and HD 61005.** The stellar core photons at r < 3 pixels radius from the center of x-ray brightness are **blue**, the halo photons at 3 < r < 6 are **red**, and the background photons at r > 6 pixels radius are colored **green**.

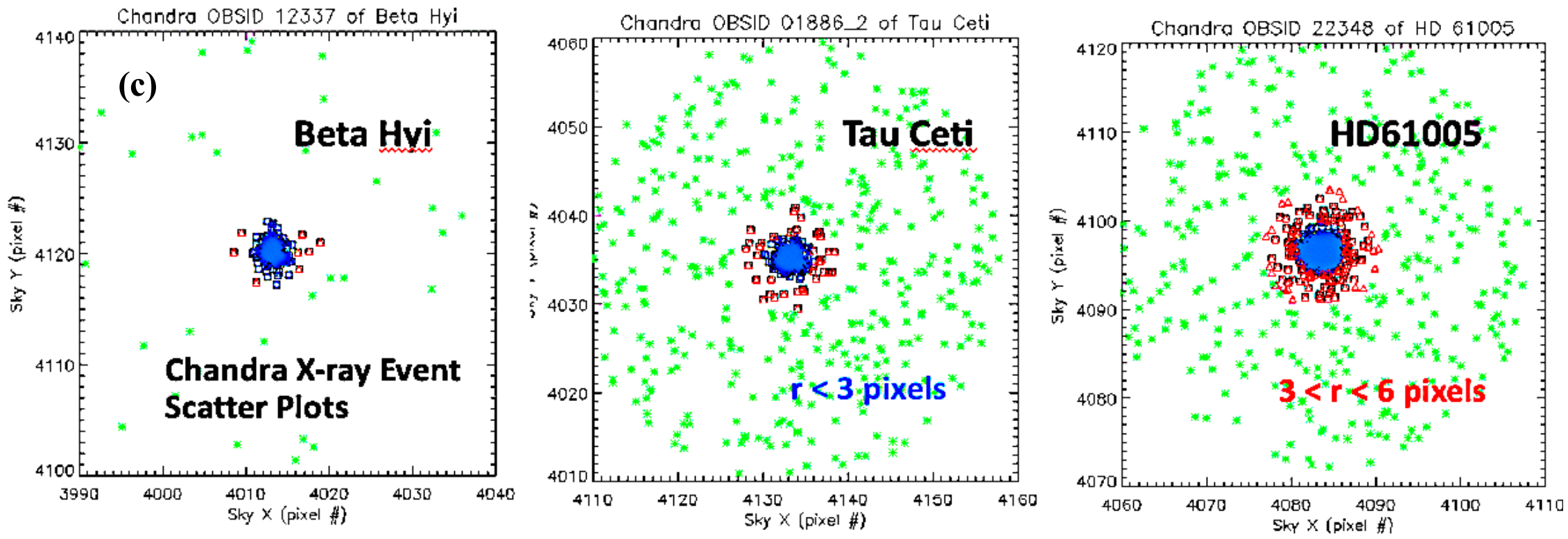

**3.1 HD 61005 System X-ray Halo Extension.** In order to test whether there is any extension of the observed Chandra ACIS-S emission beyond that expected for a point source, we performed radial aperture photometry on our measured HD 61005 image and compared it to the resulting aperture photometry of three other G-stars also observed by ACIS-S and available from the Chandra archive: Tau Ceti (G8V, ~6 Gyr, Chandra ObsID 01886), Beta Hyi (G3V, ~5 Gyr,

Chandra ObsID 12337), and 18 Sco (G2V, ~5 Gyrs, Chandra ObsID 12393; Appendix B). The results are shown in Fig 4. As can be easily seen, all three stars have a sharp core of emission going out to a radius of 2-3 pixels (~1.0-1.5") on either side of the central pixel, demonstrating the ~0.5" HWHM imaging response of the Chandra HMRA optics. HD 61005, however, demonstrates an additional halo of emission around this sharp core that extends out to at least 6 pixels around the central pixel, with ~11% (OBSID 22348) and 17% (OBSID 22349) of the total photons detected in this halo versus the sharp core (as opposed to < 4% of the detected photons being outside the sharp core of Tau Ceti and Beta Hyi). We have further tested the halo photons by re-plotting images of the three G-stars" 'sharp core" photons, i.e. all photons within 3 pixels of the stars center of brightness, versus the "halo" photons lying between 3 to 6 pixels from the object's center of brightness, vs. all photons residing at distances greater than 6 pixels in the imagery's "background". The results of this replotting are shown in Fig. 4c.

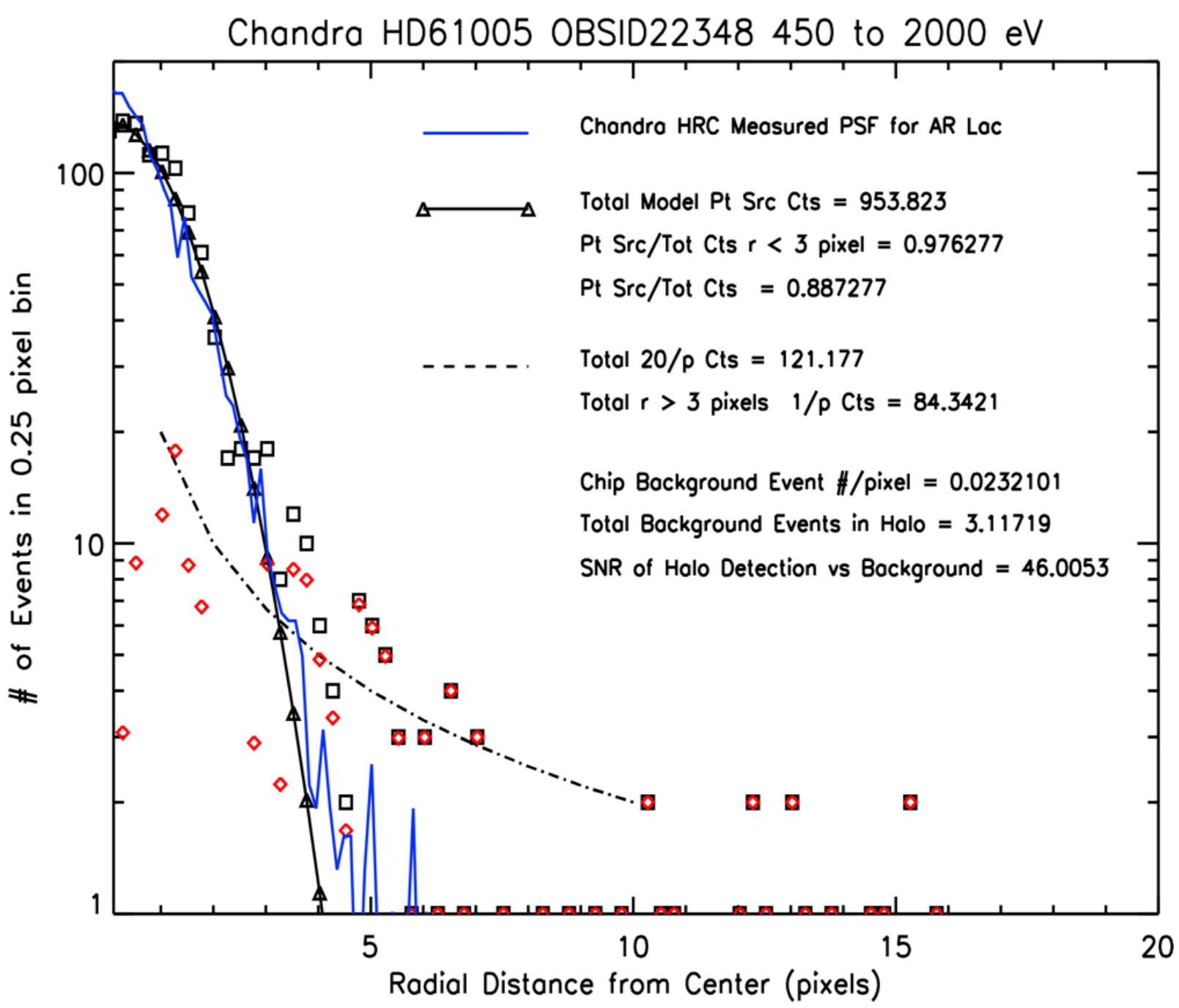

**Figure 5 – Radial distribution of the Chandra ACIS-S x-ray events detected in OBSID 22348 in 2021.** The as-observed distribution is denoted by black boxes; this distribution shows a strong central peak in the first 2 pixels but also an extended tail out to ~6 pixels before the distribution falls to the level of the background. Also shown is a best-fit point source model (triangles + solid black curve), based on Chandra long term monitoring of the AR Lac point source (blue curve). Our new halo residuals are also shown (red diamonds), and these are well-fit by a 1/projected distance law with $>45\sigma$ significance versus the background. Analysis indicates that the "core" emission extends only out to ~3 pixels, and comprises ~89% of the observed total x-ray luminosity, while 11% of $L_{x,total}$ was from the halo, and halo emission dominates at distances between 3 and 6 pixels from the center of brightness. The results for OBSID 22349 (not shown) are similar.

Another way of plotting up the x-ray event distribution is by forming a histogram plot, for an image centered on HD 61005, of the azimuthally averaged ACIS-S x-ray events in evenly spaced radial bins (Figure 5). The observed trending is consistent with an unresolved point source at the center of the imagery that extends out to ~3 pixels, consistent with the width of an ACIS-S point-source like Tau Ceti and Beta Hyi (after limiting the energy of allowed events to > 0.45 keV to avoid the effects of ongoing ACIS-S surface contamination and reacquire a sharp effective PSF: see Appendix A); the same point source structure is seen for the field object seen ~1' NW of HD 61005 (Section 2.1). Beyond that, there is an extended falloff out to at least 6 pixels distance (or a stellarcentric distance $r_h$ = 110 au assuming HD 61005 is 36.4 pc from Chandra) where the extended component signal falls below the background level (composed of instrument noise, cosmic rays, cosmic x-ray background) in the image. We cannot say if the astrosphere emission extends further, because the signal is too weak compared to the background in our imagery to tell. We can state, however, that we robustly detect a halo excess versus the background and the central point source -based on whole chip counts, the background events in the halo region r = 3 to 6 pixels total 3.1 + 1.9 (1σ) counts, while the observed signal is 90.6 counts, for a detected excess versus the background of 87 +1.9 cts, at > 45σ significance.

The slope of this extended component goes like 1/bin in the histogram, consistent with a 1/ρ (where ρ is the projected distance on the sky from the center of brightness of HD 61005) dependence for the halo emission. A 1/ρ falloff in the halo is consistent with a $n(r) \sim 1/r^2$ three-dimensional space density of emitting material produced by production from a central source of emitting material flowing out at constant velocity into a spherical cavity[5]. This matches a model scenario of unresolved direct coronal x-ray emission from the HD 61005 central host star plus emission from the entire astrosphere of the system induced by a spherically outflowing stellar wind from the host star. However, a 1/ρ dependence would also be found for emission from a spherical shell region, and there is some uncertainty in our fit of point source vs extended halo, so we cannot distinguish very well between an "entire astrosphere is lighting up in x-rays" from an outer ~55 au "astrosheath-only" emission scenario (Sections 4 and 5)[6]. It is important to note that in either case, > 80% of the total detected flux is from the central point source. While the halo x-rays are only from the extended emission source mechanism, the "core PSF" x-rays contain a small admixture of the extended emission added on top of the unresolved stellar coronal emission; for the "entire astrosphere emitting" case, this is at most ~15% of the total flux found in the core region (Fig. 5).

[5] This is a familiar case from the study of comets involving emission from a point-source like nucleus into a $10^5 - 10^6$ km surrounding diffuse coma. By Gauss's Law, $Q\,\Delta(t) = 4\pi r^2 n(r) v(r)\,\Delta(t)$ where Q is the nucleus production rate of a species, n(r) is the 3-dimensional space density of the species, v(r) is the 3-dimensional velocity of a species, and Δ(t) is a time interval of observation. Reorganizing, we have $n(r) \sim Q/[4\pi r^2 v(r)]$. Integrating along the lines of sight in a 2-dimensional image of the system, for constant outflow velocity, we find that $n(\rho) \sim Q/[2\rho v(\rho)]$, where ρ is the projected distance from the nucleus (A'Hearn *et al.* 1984). The same mathematical argument can be used to study a point-source like corona (~6 x$10^5$ km) emitting stellar wind into an extended astrosphere of radius ~ 1.5 x $10^{10}$ km.

[6] In our own solar system, the majority of charge exchange processing occurs in the region between the termination shock, where the solar wind becomes subsonic, at ~90 au, and the heliopause, where the galaxy begins, at ~120 au.

**3.2 Halo vs. Core Spectroscopy.** We tested the halo photons by plotting the spectra of HD 61005's sharp core (all events with r < 1.5" of the maximal brightness pixel) versus its extended halo emission (all events with 3.0"> r >1.5" from the max brightness pixel; Fig 6). We immediately see that: (1) there are ~10% more photons in the core + halo than the core alone, consistent with the aperture photometry curve (Fig 4a); (2) the core spectrum is continuum-dominated and peaks around 0.9 keV, hot for a mid-life main sequence G9V star, but near-normal for a young ZAMS G-star (Scelsi *et al.* 2005); (3) the halo photons range across the entire energy range, but seem most common at the lowest energies; (4) the halo spectrum is dominated by line emission; and (5) the halo spectrum looks nothing like the background spectrum.

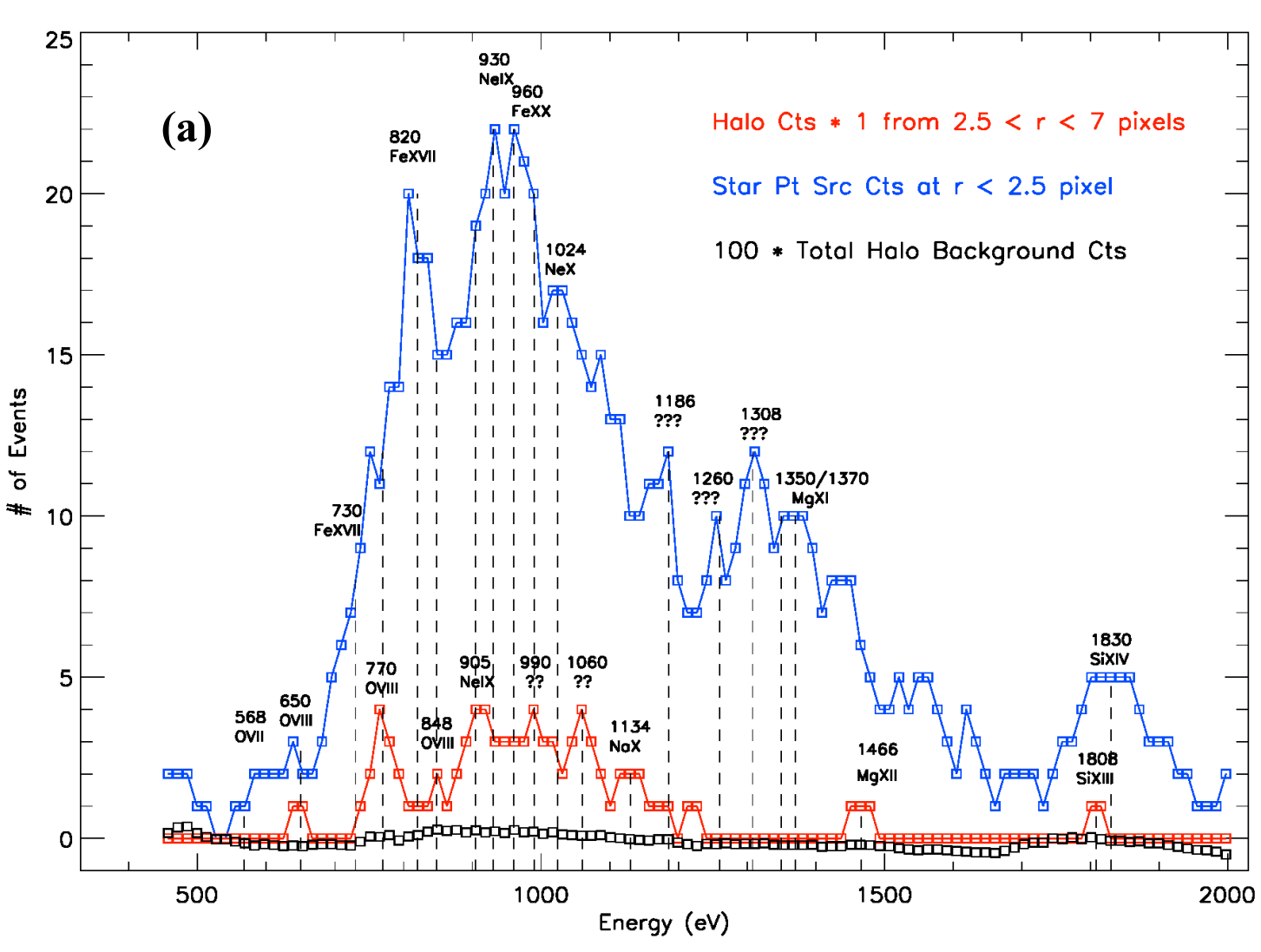

**Figure 6a – Chandra/ACIS-S 0.45-2.0 keV core, halo, and background event spectra of HD61005.** As-observed total HD 61005 spectrum broken into the two parts, core (**blue**) and halo (**red**) of the photon location image (Fig. 4c). The **blue** core spectrum is dominated by x-ray continuum, with lines of OVIII, FeXVII, FeXX, NeX, Mg XI/XII, and Si XIII superimposed (Table 1). **Black** denotes the spectrum of the background, **multiplied by a factor of 100**, that was removed from the core and halo spectra before plotting. The halo produces ~15% of the total observed events, and the background does not match any of the observed source spectral features, except for possibly the SiXIII events at 1700 – 2000 eV (potentially due to charged particle induced x-ray fluorescence in the Si-based ACIS detector).

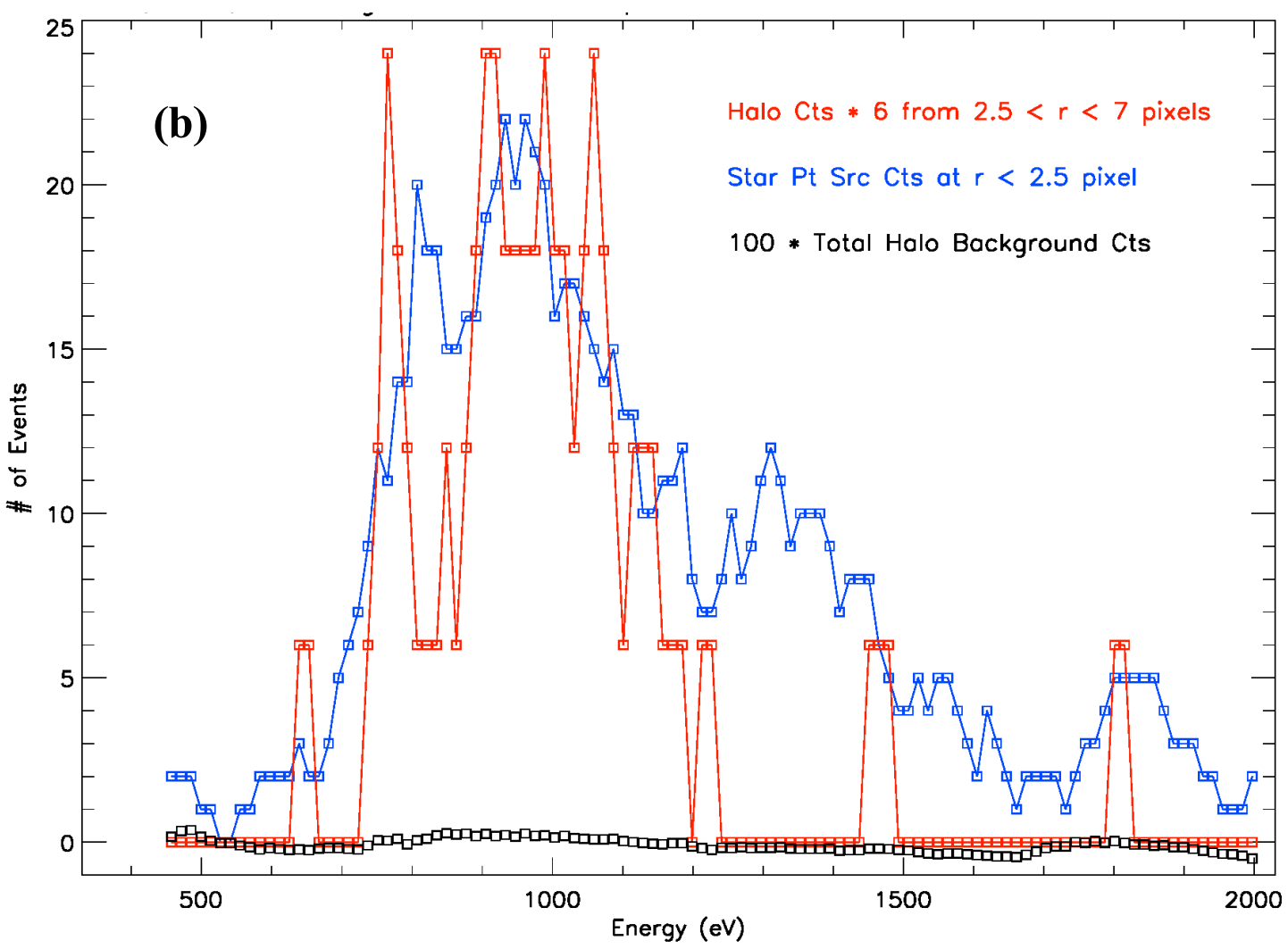

**Figure 6b -** Same as **(a)**, but with the halo spectrum multiplied by a factor of 6 to match the amplitude of the core spectrum. The halo spectrum (**red**) is dominated by emission lines, especially from OVII (568 eV), OVIII (654, 770, 848 eV), NeIX (905 eV) (Table 1). The 1050 eV line is unidentified, the line at ~1134 eV may be due to NaX CXE, and the low significance peaks at 1480 and 1800 correspond to possible hydrogenic lines of MgXII and SiXIV, respectively (see text).

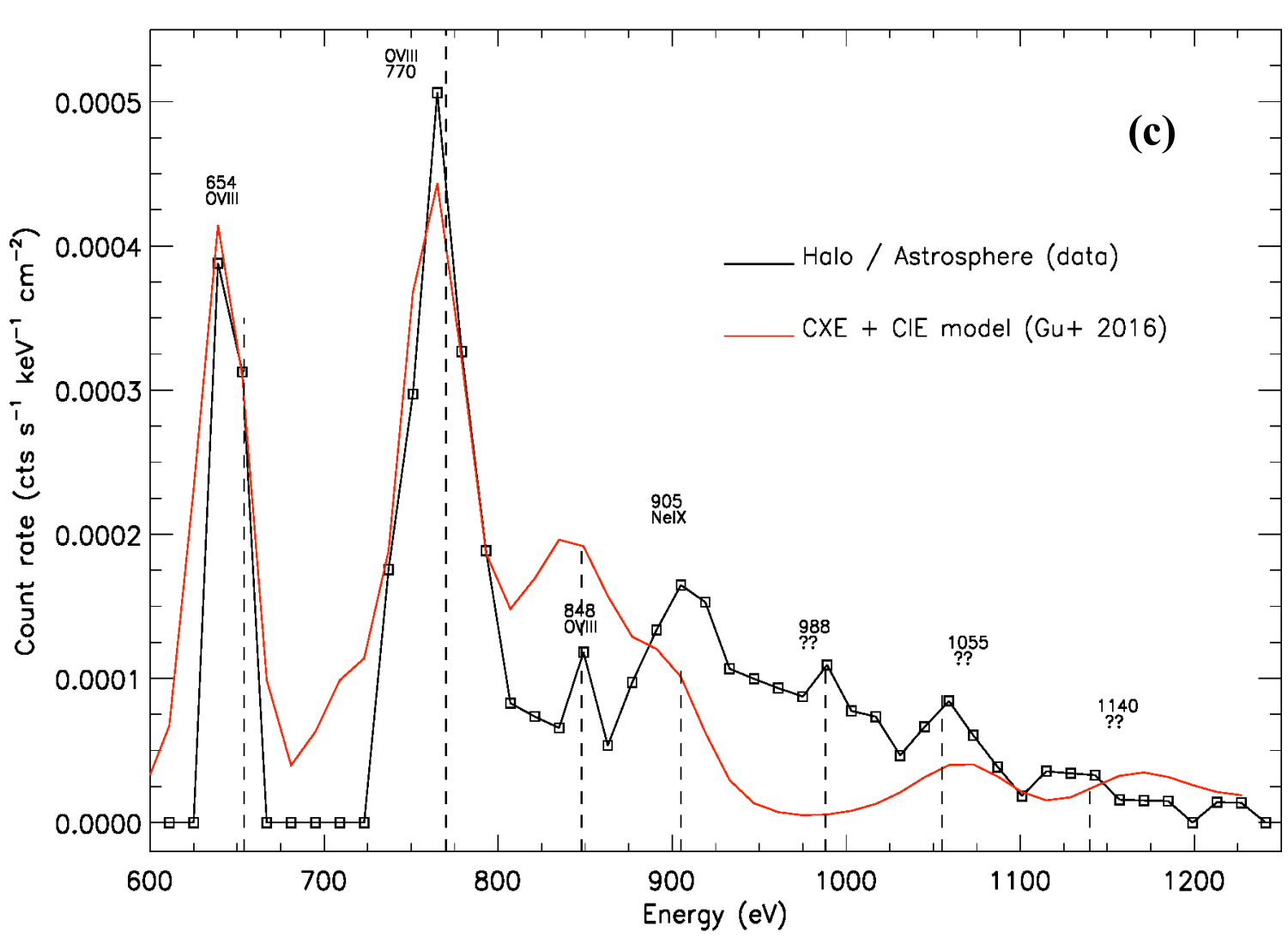

**Figure 6c - Chandra HD61005 halo flux spectrum** normalized by the 2021 ACIS-S effective area, spectral bin width, and integration time (red curve), plotted versus a model X-ray emission prediction for an ionized solar abundance plasma of 1 keV in collisional ionization equilibrium plus an 0.3 keV wind undergoing charge exchange (**black curve**; Gu *et al.* 2016). The prominent lines are due to O VIII charge exchange at 654, 770, and 848 eV and NeIX at 905 eV. For this model, emission beyond 0.9 keV is dominated by the collisional plasma. The model fit to the observations at 1-2 keV is not good; more work, beyond the scope of this first observational results paper, needs to be done modeling this spectrum for non-solar abundance, hot populations. The model fits are good enough, though, to show that the strong lines Oxygen lines found in the halo emission are those expected from CXE.

The HD 61005 "extended halo" spectrum is clearly different. A $\chi^2$–squared test comparing the total halo and core spectra returns a $\chi^2_\nu$ value of 46 per degree of freedom, for 110 spectral bins, versus their being the same type of spectrum. The same $\chi^2$–squared test comparing the total core spectra from the two different visits that were conducted returns a value of 1.5. The halo spectrum is dominated by line emission from OVIII and NeIX (Table 1), with possible contributions from OVII and NeX at much lower count rate and confidence. None of the FeXVII, Mg XI and XII or SiXIII emission seen in the primary star's (core) spectrum is apparent, and continuum contributions are very weak.[7] Along with CV/VI and NVI/NVII at 280 – 470 eV, which were filtered out of our analysis due to the ACIS-S contamination problem (Appendix A), OVII, OVIII, and NeIX are the main lines observed for solar wind CXE in our solar system (Lisse *et al.* 2001, 2005; Kharchenko *et al.* 2003, Bodewits *et al.* 2007; Mullen *et al.* 2017; Fig. 6c)[8]. The strong halo NeIX line at 905 eV is especially telling – it is typically not found in stellar spectra without an accompanying much stronger NeX line[9], but is found to be very pronounced in hot stellar SWCX. Along with the imaging evidence for spatial extension, we consider the line peaks at 654, 770, and 848 eV (O VIII charge exchange) and at 905 eV (NeIX) strong proof of the detection of extended x-ray emission due to charge exchange in HD 61005's astrosphere, and of a stellar wind flux ~74 times larger than the Sun's (using the OVIII charge exchange models of Kislyakova *et al.* 2024).

---

[7] To quote Gu & Shah 2023: "Unlike radiative recombination, charge exchange always ends up in an excited state, producing strong line emissions but zero continuum".

[8] The strongest lines of charge exchange found in solar system comet spectra, (i.e. in a mature G2V-star system with known stellar wind flux density and charge state interacting with neutral CHON atoms) are the OVII 560 eV triplet and the 665/770 eV pair of OVIII lines, with the OVII typically dominant and the OVII/OVIII ratio depending on the temperature of the solar wind (i.e. whether it is the hot, slow equatorial wind, or the cold, fast, polar wind; Beirsdorfer *et al.* 2001, 2003; Bodewits *et al.* 2003, 2007; Kharchenko *et al.* 2003; Lisse *et al.* 2001, 2004). For HD 61005 we find three strong OVIII lines, and a much weaker OVII line (Table 1), thus OVIII is clearly dominant; this makes sense because the corona of HD61005 is some 6 times hotter than the Sun's (Section 2.2). The strong halo spectrum lines of NeIX 905+912 eV, without any equivalent or stronger NeIX 922 eV line, is also very characteristic of cometary charge exchange.

**Table 1 – Relevant X-ray Lines for the Chandra/ACIS HD61005 Observations**

| Energy (apparent ACIS-S line center, eV) | Emitting Species | Coronal or CXE | Reference |
|---|---|---|---|
| 654 | OVIII | CXE (detected) | Kharchenko+2003 |
| | OVIII | Coronal (detected) | |
| 740 | FeXVII | Coronal (detected as 2 lines, 727/739 eV) | Testa 2010, Shah *et al.* 2024 |
| 770 | OVIII | CXE (detected) | Kharchenko+2003, |
| | Fe XVIII | CXE (detected) | Gu+2016 |
| 820 | FeXVII | Coronal (826 eV, detected) | Testa 2010, Gu+2016 |
| 848 | OVIII | CXE (detected) | Kharchenko+2003 |
| | Fe XVIII | CXE (detected) | Gu+2016 |
| 905 | NeIX | CXE (2 lines 905/912 eV, detected) | Kharchenko+2003 |
| 930 | NeIX | Coronal (detected) | Testa 2010, Gu+2020 |
| 960 | FeXX | Coronal (detected, likely 950 eV) | Testa 2010, Gu+2019 |
| 1024 | NeX | Coronal (detected) | Testa 2010, Gu+2020 |
| 1050 | NeX/NeIX | CXE (detected) | Rigazio+2002 |
| | Fe XXII + Fe XXIII | Coronal (not seen) | Gu+2019 |
| 1060 | ??? | CXE (detected) | This Work* |
| 1134 | NaX | CXE (detected the 1127 eV line) | This Work* |
| 1186 | Fe XXIII + Fe XXIV | Coronal (detected, likely 1174 eV line) | Testa 2010, Gu+2019 |
| 1230 | NaXI | Coronal (detected) | Philips+ 2010 |
| 1308 | ??? | Coronal (detected) | This Work* |
| 1350 | MgXI | Coronal (detected) | Porquet+2010, Testa 2010, Gu+2020 |
| | MgXI | CXE (not seen) | Ewing+2013 |
| 1370 | MgXI | Coronal (detected) | Testa 2010, Gu+2020 |
| 1466 | MgXII | CXE (detected) | This Work*, Ewing+ 2013 |
| | MgXII | Coronal (not seen) | Testa 2010, Gu+2020 |
| 1530 | AlXII | CXE (not seen) | This Work* |
| 1761 | AlXIII | CXE (not seen) | This Work* |
| 1808 | SiXIII | CXE (detected) | This Work* |
| | SiXIII | Coronal (detected) | Testa 2010, Gu+2020 |
| 1830 | SiXIII | CXE (not seen) | Ewing+2013 |
| | SiXIII | Coronal (detected the 1836 eV line) | Testa 2010, Gu+2020 |

In Table 1 we list the apparent lines detected in our Chandra ACIS spectra of HD 61005, both coronal (i.e., present in the core spectrum and arising from the hot MK plasma "atmosphere" of the central host star) and in the halo (i.e., arising from the extended volume of space extending outward from the star until ~110 au). Finding a coronal line for a given atomic species means the atoms are present in the corona in abundance, so finding a corresponding CXE line for the same species is consistent with the atoms also being emitted into the astrosphere in the stellar wind. Along with the typical O, Fe, Ne, Mg, Si lines arising from stellar coronal emission and known O, Fe, and Ne 0.6 – 1.0 keV CXE lines from solar system studies, we also find a number of interesting possible line detections at ~990, 1060, and 1134 eV in the halo spectrum and possible lines at 1406 and 1808 eV due to MgXII and SiXIII CXE.

[9] Normal Ne emission from stellar coronae is dominated by the $2^1P^1 \rightarrow 1^1S^0$ 922 eV resonance line (seen in our ACIS-S observations in the 930 eV bin of the core spectrum).

The 1134 eV line is potentially the most interesting of these five higher energy halo lines, as it is likely due to NaX. Na ions are not known to be highly abundant in the solar wind. However, they are well known to be easily created by solar wind sputtering of exposed rocky materials (e.g. the pronounced sodium tails of comets (e.g. Cremonese *et al.* 1997, Ogilvie *et al.* 1998, Schmidt 2016) and the Na atmospheres and tails of the Moon (e.g. Mendillo *et al.* 1991, Smith *et al.* 1999) and Mercury (e.g. Potter *et al.* 2002, Schmidt *et al.* 2012). It is plausible that abundant rocky circumstellar grains in the HD61005 disk, feeding into the stellar corona, could be the initial source of the sodium atoms creating this line.

It is important to note, however, that we do not detect all the known CXE lines in the 0.6-2.0 keV region of the halo spectrum. Of published studies' predictions, we do not see evidence for halo CXE lines at the 1340/1864 eV energies predicted by Ewing *et al.* 2013, and some of the ones we do see are not included in the CXE model of Gu *et al.* 2016 (Table 1). We also note that the lines at E > 1.2 keV for the halo spectrum are of low statistical significance, and that the total number of counts in the spectrum is low, ~300. So, while it is tempting to infer that the lower abundance Na, Mg, Al, and Si highly-stripped ion population in the stellar wind is affected more by a relatively collisionally thick (to CXE) interplanetary medium and so these lines are weak and the most stripped ions have been removed via CXE, it is also possible that an ~0.65 keV coronal plasma does not produce as many of these highly stripped minor ions as it does the ones of lower Z. A full-up radiative-transfer CXE spectral modeling study beyond the scope of this first results paper (including the effects of ACIS-S spectral resolution, sensitivity, and noise), is required.

## 4. Observational Results and Physical Implications.

In this section, we discuss the observational results and physical implications that "fall out" of our new Chandra observations and findings with minimal analysis and modeling assumptions. In Section 5 we present a simple model that can account for all of the Chandra observations and findings. In Section 6 we tie this modeling and the physical implications together to produce further important findings, and in Section 7 we use these results to identify promising lines of future research.

## 4.1 Observational Results.

**- HD 61005 is an x-ray bright, ($L_x = 6.8 \times 10^{29}$ erg sec$^{-1}$), very XUV/SW active young (100 +/- 50 Myr old) G9 star.** Its stellar corona is emitting detectable flux at **$L_x = 5.7 \times 10^{29}$ erg sec$^{-1}$**, or ~85% of the total luminosity, with unresolved point source morphology. The Chandra circular morphology extends out to at least 3.0" radius (6 pixels) from the center of brightness, or ~110 au at the 36.4 pc distance of HD 61005, demarcating the minimal extent of an astrosphere emitting at least **$L_x = 1.1 \times 10^{29}$ erg sec$^{-1}$**. The stellar wind (SW) flux from this star is so strong, ~74 times the Sun's, that its astrosphere, as observed by Chandra x-ray emission through its flanks, appears morphologically similar (Fig. 7) to the spherical-bubble case predicted by Parker (1965).

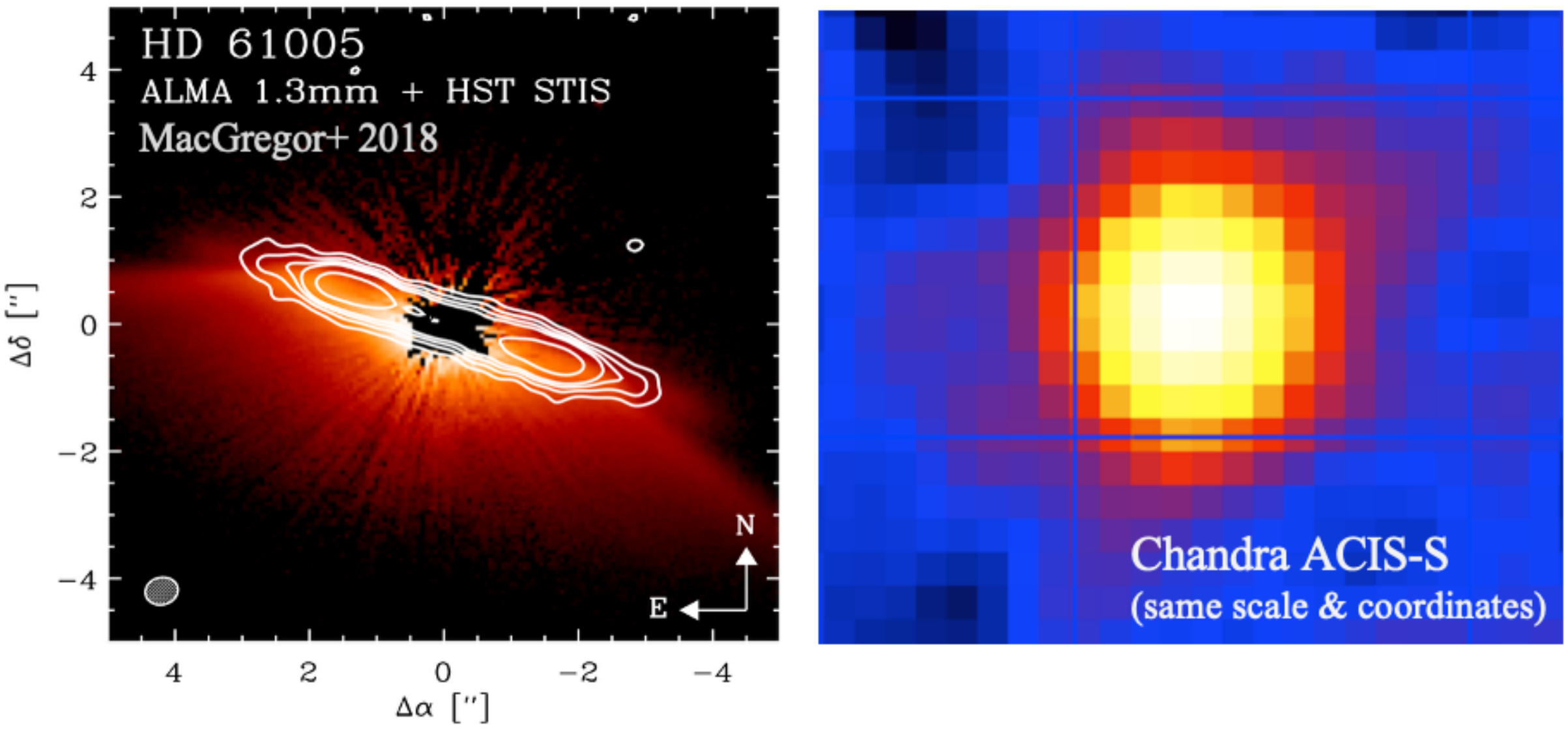

**Figure 7 – (a) *HST*/STIS (color:** Schneider *et al.* 2014**) + ALMA imagery (contours:** MacGregor *et al.* 2018**) of HD 61005,** which suggest that there are two components to the disk populated by both small micron-sized grains (HST) and larger mm-sized grains (ALMA): (1) a confined planetesimal belt between 42 and 67 au with a rising surface density gradient and (2) an extended outer halo. For scale, Voyager 1 has found the heliopause in our $L_x \sim 10^{27.3}$ solar system (Judge *et al.* 2003) at ~120 AU, and the radial distance of the transition shock in the HD61005 system we estimate below is ~55 au. **(b) Gaussian smoothed Chandra ACIS imagery of HD 61005 is an ~11 pixel diameter (= 200 au). The overall gross projected morphology is circular, with no morphological correlation to the dust features shown in (a), other than that the width of the x-ray emitting region approximately** matches the width of the ALMA large dust particle disk contours, and the X-ray emitting region appears to end roughly where the roots of the Moth's "wings" begin.

**- Immersed inside the astrosphere is a dense young disk-shaped Kuiper Belt, viewed edge-on, with abundant cm-sized dust** at $T \sim 60$ K. This provides the 60 K featureless dust blackbody SED seen by Hines *et al.* (2007), Rhee *et al.* (2007), and Schneider *et al.* (2014) with HST + Spitzer, and the ALMA linear-disk shaped morphology found by MacGregor *et al.* 2018. T = 60K dust around a G9-K0V Main sequence star ($T_{eff}$ = 5270 to 5480 K, log $L_{HD\ 61005}$ = -0.34 to -0.17 $L_{\odot}$) places it at r = 16 au (0.46") in a K0V $T_{LTE}$ = 232 K/sqrt(r) system, at $r$ = 19 au (0.54") in a G9V $T_{LTE}$ = 256 K/$r^{0.5}$ system, and at $r$ = 22 au in our $T_{LTE}$ = 282 K/$r^{0.5}$ G2V system, versus the measured ALMA belt location at 42 – 68 au (MacGregor *et al.* 2018).

**- Smaller dust particles**, produced in the disk plane either by (a) planetesimal grinding (b) SW sputtering and/or (c) ISM sputtering (outside the astrosphere boundary), **are driven out radially from the star by radiation pressure and the strong SW** (see Lisse *et al.* 2017, 2020; Chen *et al.* 2005, 2006) along the disk mid-plane. Once outside the astrosphere, these submicron grains are hit by the local ISM ram pressure and forced back into the wings of the "Moth" (Hines *et al.* 2007; Fig. 8).

\

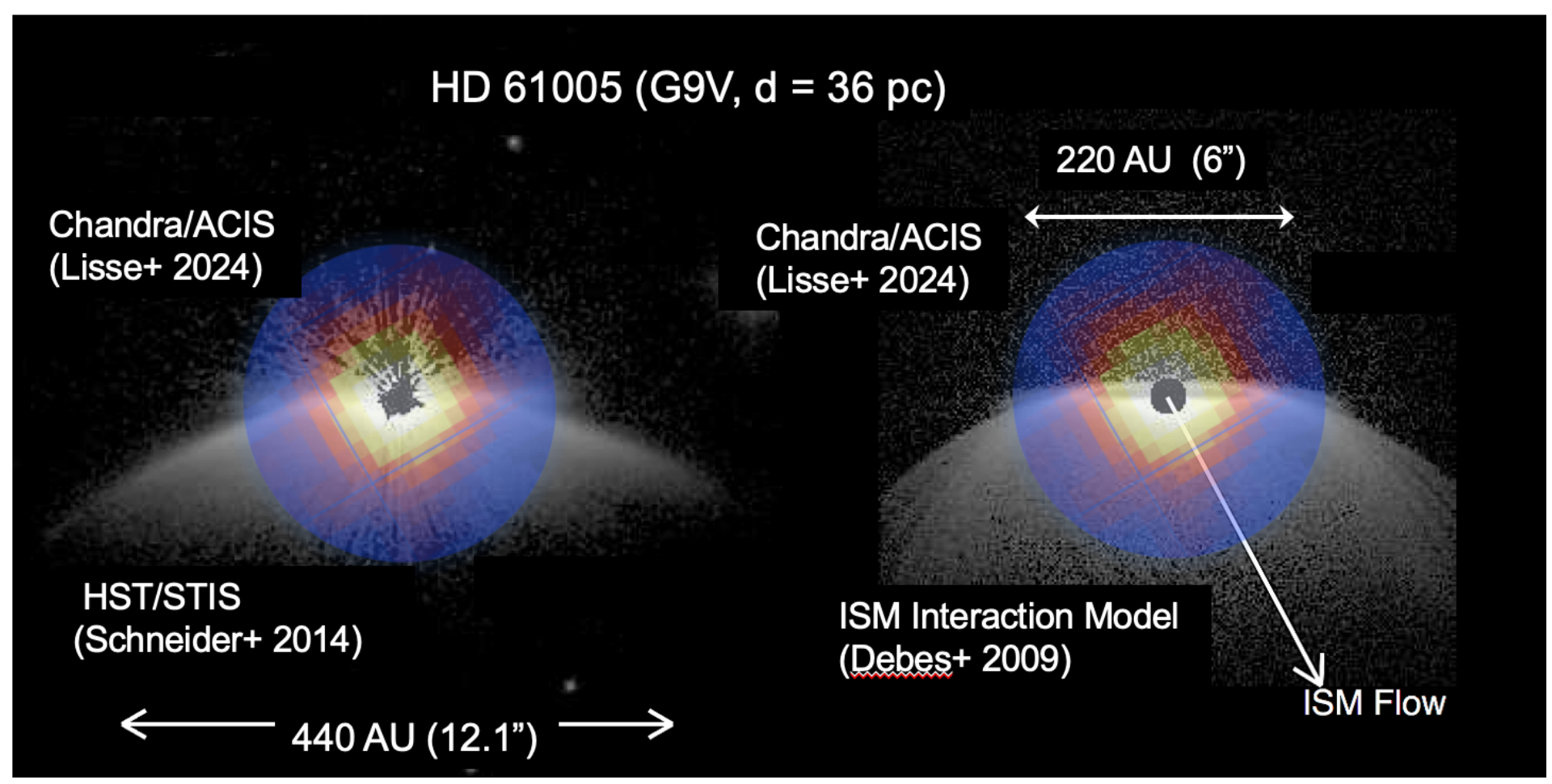

**Figure 8 – HST/STIS HD61005 imagery with our Chandra ACIS-S imaging of HD 61005 overlaid.** The Chandra x-rays extend to the base of the Moth's "Wings", in a roughly spherical pattern that is very unlike the large dust ecliptic disk + fine dust Wings. The arrow denotes the ISM flow direction determined using the complement of the host star's measured space motions (Hines *et al.* 2007). Since the system is observed almost perfectly edge-on from Earth (Hines *et al.* 2007, Debes *et al.* 2009, S. Deen priv. commun. 2024 re: TESS data showing v*sin i = v within 2%) with a velocity vector in the plane of the image, this means that HD 61005's disk is heading almost "face-on" into the VLISM.

## 4. 2 Direct Implications.

- **The x-ray flux contains a hard, point-source-like stellar coronal component with $L_x \sim 5.7 \times 10^{29}$ erg sec$^{-1}$ and an extended astrospheric "halo" component with $L_x \sim 1.3 \times 10^{29}$ erg sec$^{-1}$.** The core flux is dominated by a hot continuum, $kT \sim 0.65$ keV. Both halo and core demonstrate multiple emission lines. The bright X-ray flux is consistent with trending found for young G-stars (Ribas *et al.* 2005, Guinan & Engle 2007, Tu 2015). The multiple strong lines are consistent with a trending comparison for the continuum-removed line emission spectra of three different solar analogue stars of increasing age shows (Fig. 9), including the HD 610005 "analogue", EK Dra[10].

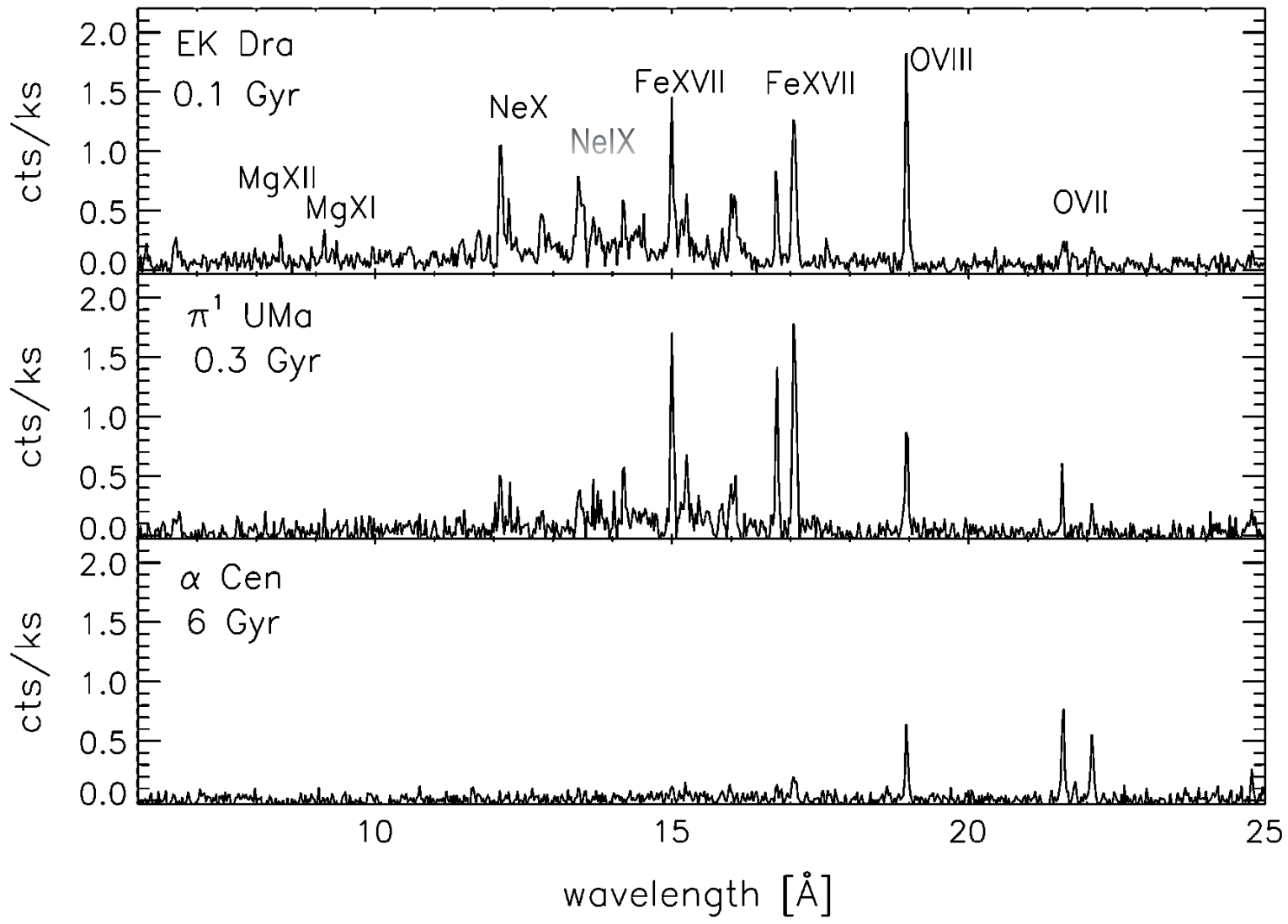

**Figure 9 – High resolution Chandra** HRC-S + LETG **x-ray continuum-subtracted spectra of 3 close solar analogue star systems: EK Dra** (age ~ 100 Myr, $L_x \sim 10^{30}$ erg sec$^{-1}$, d=34.4 pc, like HD 61005); $\pi^1$ **Uma,** age ~300 Myr); and α **Cen** (age = 4600 to 6000 Myr, like the Sun). Assuming these are all due to coronal line emission, the quick diminishment of the associated higher energy coronal line component with increasing age is readily apparent. Similar-aged EK Dra shows strong OVIII, FeXVII, and NeX corona-spheric lines as for the core of HD 61005. After Testa 2010.

- **The morphology of the system's astrosphere is roughly spherical** within the ACIS-S imaging resolution capability, suggesting an x-ray emission morphology dominated by wind outflow from the host star and not ISM magnetic fields or inflowing material. Taken together, these results support a model of HD 61005 as a highly active, fast rotating young star with a strong stellar wind.

- **The local ISM must be dense around HD 61005 for a system with ~300 times as much XUV emission and ~70 times the SW flux as the Sun's being emitted from its host star to have an astrospheric radius of only ~100 au.** Very young G and K stars with stellar wind fluxes up to 1000 times that of the Sun are common (Wood *et al.* 2005, 2021). The Sun's heliospheric extent is r ~ 120 au toward the heliospheric nose (Stone *et al.* 2019) and at least ~220 to 350 au overall based on IBEX-Lo energetic neutral atom (ENA) data (Galli *et al.* 2016, 2017), Cassini/INCA ENAS (Dialynas *et al.* 2017), IBEX-Hi ENAs (Reisenfeld *et al.* 2021), and MHD modeling (Opher

[10] Within a few 100's of Myrs, the harder lines of Mg XI/XII and NeX have greatly diminished, and by a few Gyrs' age, all but the softest OVII/OVIII line emission has disappeared. Note also that for young G-star x--ray coronal spectra (e.g., EK Dra in Fig. 9), the radiative NeX line complex at ~1024 eV is 2-3 times stronger than the ~920 eV NeIX complex; this "rule" is followed by our HD61005 core spectrum (Fig. 6a), but it is broken by the halo spectrum, where the NeIX 905 emission is clearly much stronger than the miniscule 1024 eV line flux (Section 3.2).

*et al.* 2020; Kornbleuth *et al.* 2023), so from simple scaling for the higher SW flux (Section 5) we expected HD 61005 astropsheric radii of ~ 2000 – 6000 au. No dense cloud is detected in observations of other nearby stars (e.g., HR 2882; Wood *et al.* 2005), and the Local Bubble is generally depleted in the direction of Epsilon Canus Majoris and HD 61005, but the immediate vicinity of HD 61005 is not well sampled.

**- Ram Pressure Geometry is crucial.** Our solar system's ram direction into the VLISM is at a low angle to the plane of the ecliptic - some 5° (Swaczyna *et al.* 2023), i.e., the plane of the solar system is roughly aligned with the ram direction, and the ISM is flowing into the heliosphere nearly parallel to the plane of the ecliptic (Fig. 10), so that entrained dust flows through and around the heliosphere and down its trailing "tail" (Slavin *et al.* 2012, Sterken *et al.* 2022). By contrast, HD 61005 has a large angle of attack, ~85° (Schneider *et al.* 2016; Fig. 10) versus its ecliptic, which may encourage "bleeding" of dust flowing out of the system's astrosphere into two "wings" blown back by the locally dense ISM.

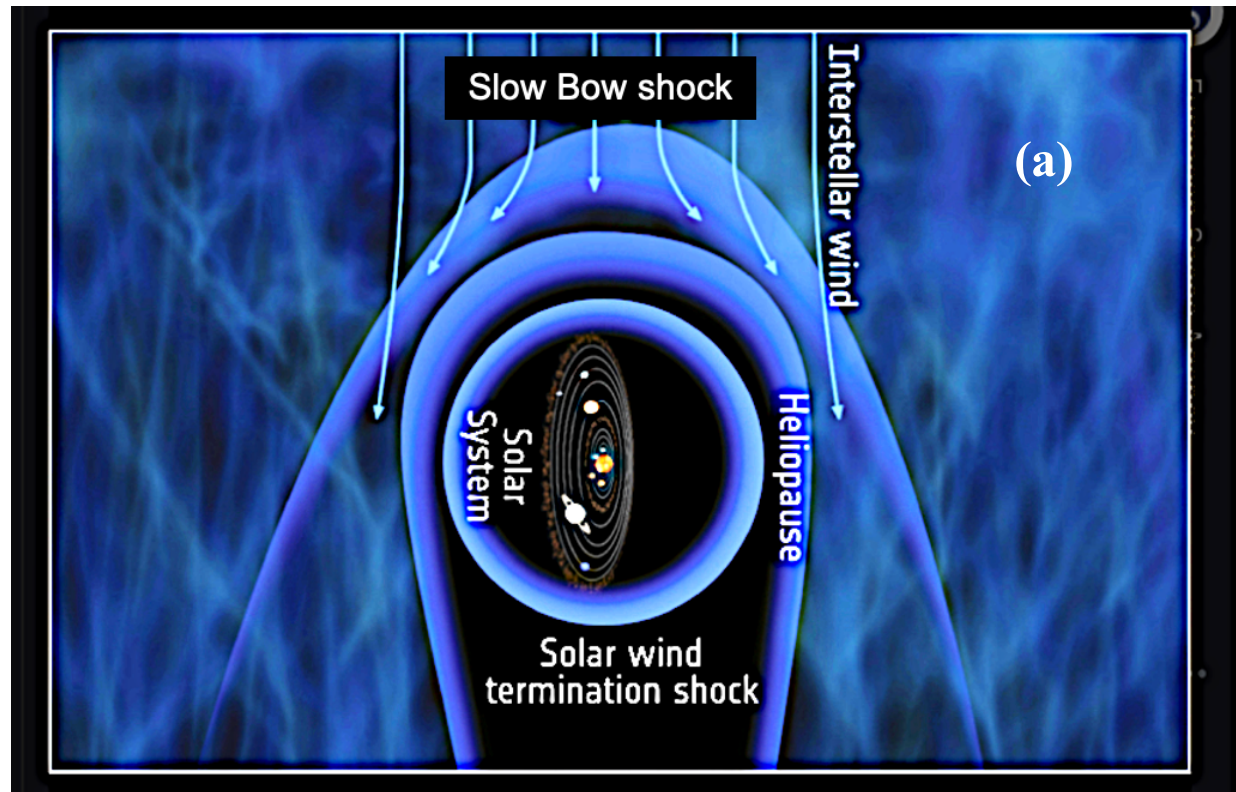

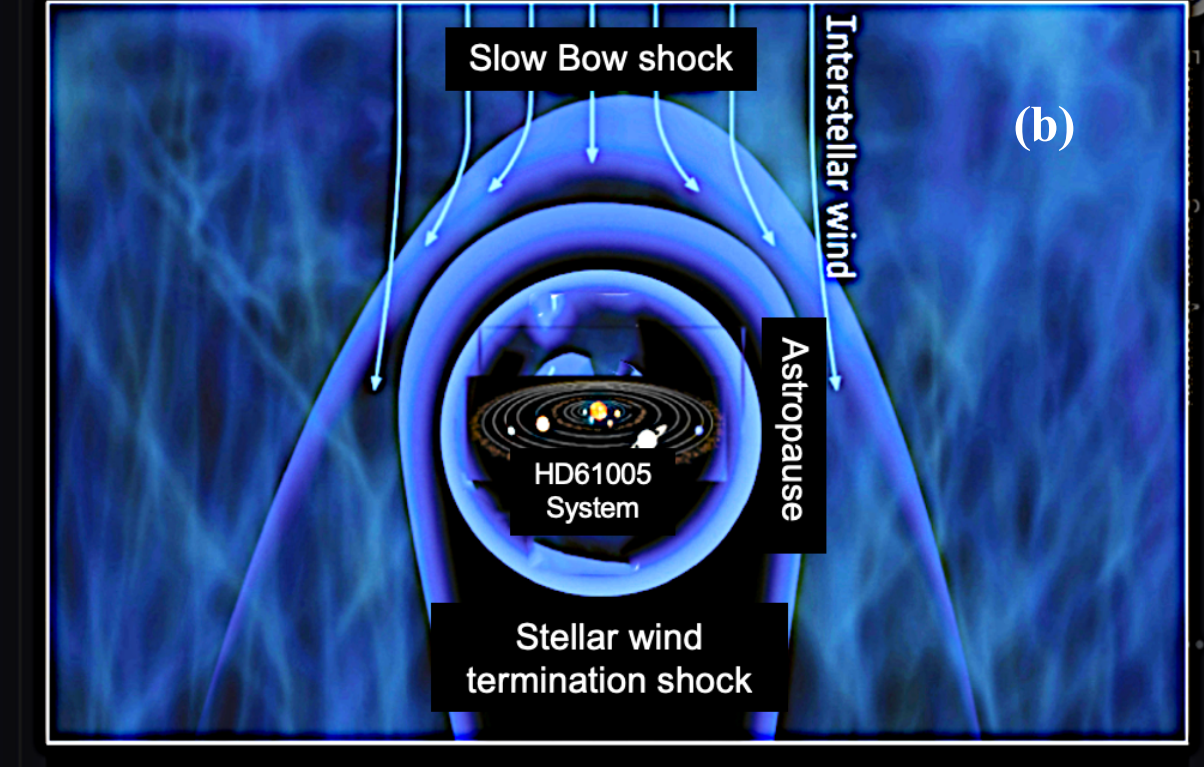

**Figure 10 – Schematic geometry for the ram direction of an astrosphere, set by the host star's motion through the ISM, and the surrounding planetary system's ecliptic (symmetry) plane.** (**a**) The situation for our solar system, with the vector from the Sun to the nose of the heliosphere (the ram direction) almost parallel (within 5°) to the plane of the ecliptic. Distance from the Sun to the heliopause is ~120 au, and from the Sun to the termination shock, where the solar wind becomes subsonic and large scale charge exchange between the solar wind and inflowing ISM material begins, is ~90 au. (**b**) The inferred case for HD 61005, with the ecliptic plane of the system ramming nearly face-on (perpendicularly) into the ISM. Assuming the HD 61005 host star's rotation pole is perpendicular to its ecliptic plane, this means that the inflowing ISM material may be "coming in the easy way" as it encounters the stellar polar magnetic field. This very different ram geometry may alter how VLISM material flows through and around the system's astropause. At the same time, the edge-on viewing geometry should greatly aid detection of the system's "wings", as they are not viewed superimposed on the rest of the circumstellar dust disk. The distance from the HD 61005 primary to the system's astropause is ~110 au, so the length scales in the two schematics is roughly the same.

**- Viewing Geometry is crucial.** Because we are observing HD 61005 edge-on, we can see the wings in cross-section as swept-back structures and not projected flat onto the sky and the ecliptic disk, appearing only as extra edges outside an ellipsoidal main disk.

- **The wings are filled with much smaller dust particles than the main parent disk**, ones that are easily accelerated by stellar wind EM forces. The mass of dust per unit scattered light flux in the Moth's wings is thus much less than in the main ecliptic disk.

- **Either the Wings are a transient phenomenon**, with a lifetime determined by the grinding/sputtering time to reduce cm-sized dust particles to micron-sized fragments, or μm-sized dust is currently being re-supplied by a Kuiper-Belt (KB) disk massive enough to be continually self-stirred, producing a steady state population of fine μm-sized dust grains. These fine grains must be small (submicron) and highly absorbing in visible light but poorly emitting in the infrared in order to maintain them at 60K at distances of 48 – 63 au form the host star (Section 4.1).

## 5. HD 61005 X-ray Physical 'Toy' Model

Our Chandra observations of the halo emission surrounding HD 61005, as dominated as they are by line emission at charge exchange wavelengths/energies, suggest a very simple toy model we can use to try and better understand the operant physics generating x-rays in the system. **The driver for this physical picture** is the need to explain the roughly spherical, but definitely extended, Chandra x-ray morphology vs the large dust planar disk + small dust "wings" morphologies (Fig. 8). Our toy model is very simple - involving a uniform inflow of ISM neutral material impinging, from the stellar polar direction, onto an astrosphere supported by the uniform spherical outflow from a very young main sequence star (Figure 10) - and is intended only as a simple order of magnitude check for the stellar wind fluxes and ISM densities required to produce the observed "flankward" x-ray emission via SWCX. The fact that the model "closes" with physically reasonable wind fluxes vs the literature is very encouraging; very enhanced (vs the galactic average) ISM densities are required, but such over-densities are not extreme vs those known to occur in nearby cold molecular clouds. It is our hope, though, that the data presented in this work will inspire future much more sophisticated HD 61005 modeling efforts allowing for 3-D structure, magnetic field pressure, MHD effects, extinction and optical depth effects, etc.

We start with

(1) $L_{x,Halo} = L_{x,CXE} = \sigma_{CXE} * n_{neutral} * [n_{sw} (n_{SW,minor\ ion}/n_{SW}) ] * v_{sw} * Volume_{interaction} * \langle E_{photon} \rangle$

(Cravens 1997, Cravens *et al.* 2000, 2001; Lisse *et al.* 2001, 2004, 2005, 2007, Wegmann *et al.* 2004). Assuming that the Sun's solar wind space density trends with heliocentric distance as $n_{SW,Sun} \sim 8\ cm^{-3}/r_h(au)^2$, that the abundance ratio of highly charged minor ions to protons is solar-system like, i.e. $(n_{SW,minor\ ion}/n_{SW}) = 10^{-3}$ (Neugebauer *et al.* 2000, Cravens *et al.* 2001, Laming *et al.* 2019), and using our new Chandra result $L_{x,Star,HD\ 61005} / L_{x,Sun} = (0.57 \times 10^{30}$ erg $sec^{-1}$ / $2 \times 10^{27}$ erg $sec^{-1}$) = 285 in conjunction with Vidotto 2021's and Kislyakova *et al.* 2024's estimate that $n_{sw} \sim (L_x/R_*^2)^{0.72}$, with $R_* \sim 0.85$, we can write $n_{sw,HD\ 61005} \sim 74 * n_{sw,Sun} = 74 * 8\ cm^{-3}/r_h(au)^2 \sim 0.049\ cm^{-3}$ at r = 110 au. We use the value of $v_{sw,HD\ 61005} = 1200$ km $sec^{-1}$ found by Pastor (2017) using data from Johnstone *et al.* (2015), about 3 times greater than the Sun's typical solar wind speed of $v_{sw,Sun} \sim 400$ km $sec^{-1}$. CXE cross sections at these velocities are on the order of $5 \times 10^{-15}\ cm^2$ (Greenwood *et al.* 2000, Gu *et al.* 2016, Gu & Shah 2023). The volume of interaction in the astropause we conservatively take from the Chandra imaging morphology as a spherical shell with outer edge at 110 au, extending inwards by 55 au (or three Chandra pixels; see Fig. 5). [If we assumed the entire 110 au radius sphere was glowing in the x-ray, then the derived $L_{x,Halo}$ estimates given below would increase by only a factor of $(110)^3/(110^3-55^3) = 1.14$]. Using $<E_{photon,halo}>$ = 0.9 keV = $1.4 \times 10^{-9}$ erg from inspection of the halo x-ray spectrum of Fig 6, this leaves us with $n_{neutral}$ as the only unknown quantity to determine in the equation for $L_{x,Halo}$.

The neutral density depends on what the stellar wind is charge exchanging with. At 50-150 Myr age, any primordial gas filling HD 61005's astrosphere (i.e., its interplanetary space) is long gone (Mamajek 2009; Fedele *et al.* 2010; Ribas *et al.* 2015; Meng *et al.* 2017). We can rule out planetary atmosphere driven CXE, as the neutral atmospheres of planets extend only out to a few planetary radii and thus involve orders of magnitude smaller total x-ray production than for even 1 single comet. Cometary x-ray emission in our own solar system has shown total x-ray luminosities of $10^{13}$-$10^{16}$ erg $sec^{-1}$; to produce the $L_x \sim 1.1 \times 10^{29}$ erg $sec^{-1}$ total x-ray emission seen in the halo, at least $1 \times 10^{13}$ comets, each massing from $10^{11}$ to $10^{14}$ kg (for a total mass of $10^{24}$ to $10^{27}$ kg) would have to have been actively outgassing in HD 61005's astrosphere while Chandra was observing it. This represents at least a Mars' mass worth of comets – so unless we were observing HD 61005 at an extraordinary time, like our solar system's purported Giant Planet Instability epoch (Tsiganis *et al.* 2005, Nesvorný & Morbidelli 2012), comets are highly unlikely to be the source of the available neutral atoms. A final point against any exosystem related source of neutrals is that the roughly spherical x-ray emission morphology found by ACIS-S for the halo component is nothing like the pronounced edge-on ecliptic plane of the system resolved by ALMA (MacGregor *et al.* 2018).

This leaves us with inflowing ISM gas as the primary source of the neutrals. Given the strong stellar wind flux expected for an $L_x \sim 0.6 \times 10^{30}$ erg sec$^{-1}$, ~100 Myr G9V Star rotating every 5 days, this would produce a roughly spherical emission geometry for a stellar wind outflow matching the Sun's (McComas *et al.* 2001) and a system where the star's polar regions are pointing towards the system's ram direction through the local ISM. An interaction involving charge exchanging with inflowing ISM neutrals can also easily have a large power output because it can involve the entire stellar wind minor ion production of the star over many au's of distance. Further, it allows us to suppose that most of the x-ray emission is occurring in the astropause boundary region, where the local ISM neutrals are first encountering the outflowing stellar wind, and to use simple fluid pressure balance to estimate the local ISM density:

(2) $$n_{SW} * v_{SW}^2 = n_{ISM} * v_{ISM}^2$$

where we have assumed that dynamic pressure components are dominant and have ignored, to first order, any magnetic pressure contributions as unimportant (because of the intense stellar wind flux [~74 times solar] and the low estimated ISM magnetic field strength of ~3 μGauss local to HD 61005 [Frisch *et al.* 2022]), and the low ISM thermal pressure. Dominance of the stellar wind pressure is also reflected in the spherical x-ray morphology of the emission (Figs 4, 7 and 8; Parker 1965).

Plugging in the values $n_{SW} = 0.049$ cm$^{-3}$ at ~110 au and the $v_{SW} = 1200$ km/s adopted above, using $v_{ISM} = 25$ km sec$^{-1}$ (Debes *et al.* 2009), one finds that $n_{ISM} \sim 110$ cm$^{-3}$ (99% HI and $H_2$) surrounding HD 61005. Compared to the ~0.2 cm$^{-3}$ density of the Very Local Interstellar Medium (VLISM; Swaczyna *et al.* 2020) surrounding our Sun, this is a very large value for the density of ISM neutrals, but not unheard of in the outer regions of the Local Bubble (Redfield & Linsky 2008, Frisch *et al.* 2011, Opher & Loeb 2024); it is reasonable in light of the system's known, pronounced dust wings demonstrating the strong effect of inflowing ISM material on its circumstellar dust (Debes *et al.* 2009 required a neutral density of ~10$^2$ cm$^{-3}$ or greater to create the wings of the system using dynamical ISM ram pressure.). It is also not unreasonable considering that simple scaling from our solar system's heliopause at ~120 au, as determined by Voyager 1 & 2 measurements (Krimigis *et al.* 2019) for an $L_x \sim 10^{27.3}$ star would put HD 61005's astropause, for a stellar wind density falling off as $1/r^2$ distance from the star and $n_{sw_HD61005} \sim 74 * n_{sw_Sun}$, at $(74)^{0.5} * 120$ au = 1030 au, all other things being equal. Thus the ~110 au, greatly contracted extent for the size of HD 61005's astrosphere we find with Chandra is another hallmark of the dense local ISM's effects.[11]

According to Redfield & Linsky 2008, HD 61005 likely lies in the nearby Blue Cloud contained in the Local Interstellar Bubble, or an extension of the G or Cet Clouds. None of these have shown the $N_H \sim 10^{21}$ H cm$^{-3}$ density values in their line of sight surveys predicted by our toy model (110 cm$^{-3}$ * 36.5 pc * 3.1 x $10^{18}$ cm/pc = 1.3 x $10^{22}$ cm$^{-2}$); the largest $N_H$ values seen in their survey are ~$10^{19}$/cm$^2$, but they also did not measure the known extremely high density Leo Cloud complex located at distances between 10 and 25 pc of Earth with $N_H$ up to ~3000 cm$^{-3}$. Their sparse sky sampling's best LOS $N_H$ measurement appropriate for estimating the ISM density at HD 61005 was for the G4V star HR 2882, 5 deg away on the sky (at l = 250.5, b = -8.97 from l = 246.38, -05.63 HD 61005) and at d = 21.8 pc, 14.7 pc closer to us than HD 61005. According to Wood *et al.* 2005, the column towards HR 2882 has log $N_H$ (cm$^{-2}$) = 18.54; when scaled linearly to 36.5 pc, this becomes 18.54 + log(36.5/21.8) = 18.8.

If Wood *et al.* (2005)'s ISM material is uniformly distributed along a 36.5 pc column, it must have average space density (6.3 x $10^{18}$/cm$^2$)/1.2 x $10^{20}$cm = 5.1 x $10^{-2}$ cm$^{-3}$, more than 3 orders of magnitude less than the 110 cm$^{-3}$ of our toy model. If instead the toy model's 110 cm$^{-3}$ is in a column local to HD 61005, the column length needs to be smaller than $d_{local}$ = (6.3 x $10^{18}$/cm$^2$)/110 cm$^3$ = 5.3 x $10^{16}$ cm = 0.016 pc = 3360 au to be consistent with the HR 2882 ISM measurement. Another possibility, of course, is that there is abundant ISM cloud density in the direction of HD 61005/HR2882 located behind HR2882, between 21.8 and 36.5 pc, that has not yet been measured. The derived hydrogen column of $N_H \sim 1.7$ x $10^{20}$ cm$^{-2}$ from our APEC modeling (Section 2) is an order of magnitude higher, suggesting that there is an ISM density enhancement along the line of sight. ***It thus seems safe to assume that if HD 61005 lies in a region of enhanced ISM density, it is localized to between 20 and 40 pc from Earth.*** It also strongly suggests that a future high resolution Lyman-$\alpha$ line-of-sight study, one that accounts for significant stellar wind mass loss and charge exchange, be made towards HD 61005.[12]

[11]Note that the 55 au emitting region thickness we measure with Chandra is about twice the ~30 au heliosheath thickness revealed by Voyagers (Krimigis *et al.* 2019, Stone *et al.* 2019). This increased sheath thickness may be due to the highly compressed nature of the HD 61005 astrosphere – something that future detailed modeling of HD 61005's astrosphere, beyond the scope of this initial observational results paper, should investigate.

Putting the final $n_{ISM}$ piece of the puzzle into the equation for $L_{x,Halo}$ we obtain for a 55 au wide (3 Chandra pixels) CXE interaction region extending from a transition shock located at 55 au out to an astropause at 110 au the x-ray luminosity

$$L_{x,Halo} = 5.0 \times 10^{-15}\ cm^2 * (110\ cm^{-3}) * [0.049\ cm^{-3} * (10^{-3}\ \text{minor ions/proton}) * (1200 \times 10^5 cm\ sec^{-1}) * [\ 4\pi/3\ (110*1.5 \times 10^{13} cm)^3 - (55*1.5 \times 10^{13})^3\ ] * 1.4 \times 10^{-9}\ erg$$

$$= 0.75 \times 10^{29}\ erg\ sec^{-1}$$

The factor of unity consistency between this back of the envelope CXE luminosity estimate and our observed Chandra ACIS-S luminosity[13] $L_{x,Halo} = 1.3 \times 10^{29}$ erg sec$^{-1}$ allows us to state that a mechanism involving CXE near an astropause at ~110 au from the host star, occurring between a strong stellar wind outflow and a dense local ISM, can very plausibly explain the Chandra observations of the system taken in February 2021.

Improved agreement between the toy model prediction and observation could be found by better determining f, the minor ions/proton ratio, likely significantly higher than $10^{-3}$ for a wind emitted by a corona 6 times hotter than the Sun's; by better measuring the stellar wind outflow velocity; by better understanding the pressure balancing between two diffuse astrophysical fluids; and by performing detailed calculations for each of the expected charge exchange reactions producing the halo line emission (all beyond the scope of this first results paper). When we consider that the reported Chandra ACIS-S $L_{x,Hal}$ is a lower limit due to the sensitivity limitations imposed by the sky background flux (likely a 10-20% effect) and by our filtering out all low energy (~0.1 – 0.4 keV photons, likely decreasing the current halo flux by up to 50%), we realize that revisiting HD61005 using much longer visits with an x-ray detection system sensitive down to 0.1 keV, like

[12] Unfortunately HD 61005 lies just outside the survey fields of the radio survey of dense galactic clouds performed by Dame *et al.* 2001 & 2011, so we cannot utilize this extensive survey to resolve the issue.

[13] Actually a lower limit to the true halo CXE luminosity, since the halo can extend out farther than the Chandra background lets us measure, and we know we are insensate to any 0.1 0.45 keV luminosity.

XMM, is highly desirable. [We note here again that Chandra's ability to detect the purported CXE emission versus the background depended not only on the ~74x stronger stellar wind flux, but also the ~570x larger local ISM neutral density and the 2.45x larger CXE cross sections than prevail in the solar system (Wargelin & Drake 2001, Kislyakova *et al.* 2024).]

## 6. Discussion.

If our overarching physical picture is correct, then HD 61005 would be the first main sequence Sun-like, G-star exosystem with a resolved astrosphere (all the other known, imaged astrospheres have been imaged have around either O/B/A or AGB host stars (Chick *et al.* 2020); e.g. B7-8III Delta Ceph (Marengo *et al.* 2010), or B1I star Kappa Cas (HD 2905; Katushkina *et al.* 2018).

Pioneering global heliosphere modeling (Baranov & Malama 1993) showed that in our heliosphere, the charge exchange between the solar wind and incoming LISM neutral hydrogen results in the location of the termination shock at ~90 au and heliopause at ~150 au. One would thus expect that for the HD 61005 star that has the stellar wind flux 74x larger than the solar wind, the wind termination shock and the astropause would be located much further from the star than the 110 au we observed with Chandra. Such a close location of an astropause interface strongly points to the crucial role of charge-exchange between stellar wind ions and ISM neutrals in controlling the overall structure and size of an astrosphere.

This latter point is very important for answering the question "Why don't all young ZAMS stars show strong resolvable X-ray astrospheres like HD 61005's?", since we can expect all young, late-type stars to generate very strong stellar winds early in their lifetimes. These stars have to be lying in a surrounding region of highly enhanced neutral density as well, and by the time they are on the main sequence, the required density enhancement can only be found via fortuitous location in one of the galaxy's dense interstellar clouds.[14]

[14] We note that while there may be good independent agreement between the order of magnitude neutral density in denser clouds of the ISM and the neutral ISM density we estimate with our CXE x-ray generation toy model, there is some tension between our HD 61005 LISM 110 $cm^{-3}$ neutral density number and the ~$10^0$ $cm^{-3}$ modeled by Maness *et al.* (2009). We note this discrepancy here with the hope that it can be resolved in the near future using Lyman alpha absorption measurements along a direct line of sight to HD61005.

The combination of a compact, high stellar wind flux impacting a dense local medium is why HD 61005's astrosphere is detectable from Earth. The special combination of a close G-star observed edge-on transiting through a dense molecular cloud means that the astropause for a system with $\sim 10^2$ times as much stellar wind is still roughly at the same astrocentric distance as that of our own solar system, but with $\sim 10^5$ times as much X-ray production per unit astrosphere volume as in the solar system. Turning the problem around, we can state that the reason that CXE from main sequence exo-astrospheres has not been resolved until recently is that with current instrumentation, it required a nearby system with $\sim 10^5$ times the CXE X-ray production of the solar system for extended emission to be detected. A case in point are the Chandra ACIS observations of the Alpha Cen system (Ayres 2009, 2023) – no astrospheric signal for this system has ever been detected, despite this system being only 1.3 pc distant, containing a close solar analogue G2V star (plus a close binary partner K2V), of similar age to the Sun, and observed with a viewing geometry within 11 deg of being edge on (Akeson *et al.* 2021). The stellar wind flux * VLISM neutral density flux product for Alpha Cen is just too small to produce detectable CXE X-ray emission from just a few light years away.

It is tempting to posit that the observed X-ray emission must be optically thin in the astrosphere because there is no overt asymmetry with respect to the ram direction through the ISM, i.e., the observed emission morphology is not appreciably much brighter in the upstream direction of the disk-ISM interaction, suggesting that the optical depth is low and the supply of interstellar neutrals flowing into the system at $v \sim 30$ km sec$^{-1}$ is not being depleted as they cross through the HD 61005 astrosphere. **However**, since a model of a strong, spherically symmetric host stellar wind blowing a spherical cavity in the ISM at $v \sim 1600$ km sec$^{-1}$ (Johnstone *et al.* 2015) and charge exchanging mainly in a thick shell near its astropause is also consistent with the data, we cannot currently say much more about the optical depth or the radial extent of the charge-exchange-emission region.

Concerning the outermost radial extent of the HD 61005 astrosphere and the thickness of its outermost astropause, our X-ray measurements can only tell us the minimum total radial extent observable above the instrument and sky backgrounds. Since we know the astrosphere does not extend to infinity, the supply of highly-stripped minor ions must run out somewhere nearby. A useful calculation to get a sense of the astropause's thickness is to estimate the total number of highly-stripped minor ions in the astrosphere and compare it to the required volume of the nearby

ISM to provide the same number of neutrals. From the toy model density estimates of Section 5, we have that the total $n_{sw}$ in an 110 au radius HD 61005 astrosphere is ~ 9 x $10^{45}$ ions. The total neutral density in the nearby ISM was determined to be in the 110 $cm^{-3}$ range, implying that an ISM sphere of radius $[(9 \times 10^{45}/(110\ cm^{-3} * 4/3\pi)]^{1/3}$ / (1.5 x$10^{13}$ cm/au) = 18 au, or about 16% of HD61005's observed ~110 au x-ray astrosphere extent, would contain about the same number of neutrals. Another useful way to view this is to realize that while HD61005's stellar wind is very strong compared to the Sun's, emitting ~ 70 * 2.5 x $10^{-14}$ $M_{sun}$/yr * 2x$10^{30}$ kg/3.1x$10^{7}$sec/yr = 1.1 x $10^{11}$ kg/sec of ions. By contrast, the ISM mass inflow rate is on the order of 3.14 * (110 au* 1.5 x$10^{13}$ cm/au)$^2$ * 25 km/sec * 1 x $10^5$cm/km * 110/$cm^{-3}$ * 3 x $10^{-27}$ kg/particle= 4.5 x $10^{12}$ kg/sec of neutrals, about 40 times higher. There are more than enough ISM neutrals to power charge exchange x-ray emission in this system, and only a few % of the impinging neutrals are ionized as they fly through the system, explaining why there is no obvious ram direction vs anti-ram direction x-ray brightness asymmetry in our Chandra imaging (Figs. 3, 7, and 8).

Finally, we note that "Mothian" behavior should be common in young systems, and all young G-stars with fast, host-star rotations, strong stellar winds, and young and dynamically hot circumstellar dust disks should show disk with associated "dust wings". Our ability to detect it, however, depends on the sensitivity of our X-ray telescope and the surface brightness of the astrosphere, which can vary by many orders of magnitude depending on the exosystem's proximity (as ~1/distance$^2$), age (in a roughly logarithmic relationship; Ribas *et al.* 2005, Guinan & Engle 2007, Johnstone *et al.* 2021), and interacting local ISM density (as density$^{3/2}$, since increasing the local ISM density both increases the total amount of astropause emission, as well as concentrates it by contracting the astrosphere's radius). Of the handful of < $10^8$ Myr old nearby disk hosting G-stars known (~15; Rodriguez & Zuckerman 2012), one of the best candidate systems for this kind of study at age $30^{+10/-20}$ Myrs and distance = 43 pc is the G7V star HD 202917 (Schneider *et al.* 2016). Although at 30 Myrs age it is probably still technically a pre-main sequence star, the image reduction results of Schneider *et al.* (2016) argue strongly for a central, nearly edge-on disk sporting bowed-back wings at large radial distances. Another good young, Sun-like candidate for future study is the G2V system HD 107146 (80-200 Myr and $d$ = 29.5 pc; Ardila *et al.* 2004, Williams *et al.* 2004, Corder *et al.* 2009, Marino *et al.* 2018). Brighter, better studied, and of a more mature early main sequence age than HD 202917, this system is unfortunately observed almost face-on, so any wing structure is likely projected onto the sky and thus harder to interpret.

## 7. Open Questions, and Implications for Future Work

- Our Sun was likely in a similar state at one point in its young life, when it was ~$10^8$ yrs old and transiting through a dense part of the ISM. The new results presented here should be adjusted for the late G-star (G9V) nature of HD 61005, and compared in detail to the latest "Sun in Time" project findings (Ribas *et al.* 2005, Guinan & Engle 2007, Johnstone *et al.* 2021). We can also ask if HD 61005's observed dust wings are co-spatial with the twin swept-back arms of many current solar system heliosphere models (e.g., Opher *et al.* 2015, 2020, 2021; Fig. 8)[15].

- These new Chandra measurements will be valuable for testing the results of solar heliosphere models adapted for the thicker, faster wind of a young G-star of solar abundance (but of very similar extent, ~110 au from host star to astropause for HD 61005 and ~120 au from host star to heliopause for the solar system), especially after measuring the local VLISM hydrogen density using Lyman alpha absorption line measurements towards HD 61005 and allowing for the 90° difference in ecliptic plane - ram direction angle. Thus HD 61005 should be re-observed in detail by future X-ray instrumentation with better sensitivity and image resolution in order to accurately define the astrosphere's outer boundary. Early (pre-2008) archival GALEX data should be re-analyzed and modeled in detail, since for every stellar-wind minor ion that charge exchanges, approximately 1000 protons and 100 alpha particles charge exchange as well. There should be strong extended emission in Lyman-$\alpha$ and in the He 304Å line detected as well, with their own emission measures - and in fact HD 61005 is found as a bright GALEX source, with a possible diffuse extension above the 4-5" wide PSF of the GALEX beam (http://www.galex.caltech.edu /researcher/techdoc-ch5.html).

- How do HD 61005's pronounced dust "wings" fit into all this, other than as direct evidence for strong ISM-system interactions? The observed X-ray emission morphology extends to approximately the base of the wing roots (Figs. 7 and 8). While the roughly spherical X-ray emission does not track the strong edge-on ecliptic disk + wing dust structures, arguing that dust is not involved in the X-ray generation, the converse is not necessarily true - the dust wings seem to start where the X-ray emission ends. But why should the dust care about the x-ray emission? The simplest answer is that the dust exterior to the X-ray emission is also exterior to the system's astrosphere, as demarcated by the observed halo emission. It is thus unprotected from the influence of the local ISM, and can be swept back by the ram pressure induced by the ~ 25 km/sec (Hines *et al.* 2007, Debes *et al.* 2007, Maness *et al.* 2009; Section 1) between the two. The question of whether erosive processes in the dense edge-on, mm-sized dust particles disk are creating large amounts of smaller micron-sized dust particles that are then blown out by radiation pressure through the astropause, and thence into the wings, or whether the dust disk sticks out past the astropause directly into the incoming ISM flow, where its large particles are sputtered and the resulting micron-sized fragments are swept into the wings is still open. (Hypothesizing that there are HD 61005 system planetesimals outside the system's astrospheres is entirely plausible, in light that our own solar system's Oort Cloud comets spend > 99 % of their orbital lifetimes outside our heliosphere). Another possibility is that we are seeing instreaming ISM dust being focused and diverted around the astrosphere and into the wings, and this may be aided by the system's ram direction being "face-on to the ecliptic, parallel to the host star's polar axis" orientation. We require better imagery than we currently have from Chandra/ACIS measurements to search for subtle structures – like an ISM ram/anti-ram asymmetry expected from focusing of instreaming ISM material – to distinguish between possible causes. We may have to wait until the next generation of improved sensitivity X-ray imaging spacecraft, like the proposed AXIS mission (Reynolds *et al.* 2023), to search for this effect.

[15] Testing, refining, and improving models of our astrosphere is also of great topical interest for the NASA New Horizons mission currently traveling through our outer astrosphere and for the proposed NASA Interstellar Probe mission (Brandt *et al.* 2023).

- The astropause distance of HD 61005 should vary with stellar activity rate and local ISM density, and is likely highly compressed versus its more usual ~1000 au extent (Section 5). This small astropause distance, coupled with near-maximal local ISM density and high stellar-wind flux, has strong implications for surface processing of planetesimals currently extant in the HD 61005 system (see Opher & Loeb 2024). It also has implications for our own solar system's development - our Sun was in a similar state at times in its young life, when it was ~$10^8$ yrs old (Guinan & Engle 2007), and transiting through a dense part of the ISM such as a Giant Molecular Cloud (Stern 2003, Opher & Loeb 2024) during its ~250 million year orbit around the Galactic Center.

## 8. Conclusions.

- Comparing our Chandra observations of HD 61005 G9V (very young, 40-130 Myr, 36.4 pc) to archival Tau Ceti measurements (old, 6-7 Gyr G8V, 3.7 pc) is proper, as both are nearby, of roughly solar metallicity, and harbor similar kinds of dust disks. Upon comparison, we find HD 61005's X-ray emission to be hyper-extended, with a pronounced "halo" of low-level emission not found for point-source-like like Tau Ceti.

- The X-ray extension morphology does NOT appear to follow that of the well-known HD 61005 dust disk + wings. This implies that the x-ray morphology is dominated by the structure of the system's strong, roughly isotropic stellar wind interacting with the VLISM.

- The achieved Cycle 22 ACIS-S Count rate of 0.032 cps at 36 pc distance from 0.45 – 2.0 keV equates to an X-ray luminosity of ~6 x $10^{29}$ erg $sec^{-1}$. This is ~1500× the Chanda archival luminosity for G2V 18 Sco, ~5000× the Chandra archival luminosity for G8V Tau Ceti, and 4000× the archival luminosity of ~1.5 x $10^{26}$ erg $sec^{-1}$ for G1V Beta Hyi. On the other hand, the observed high $L_x$ is consistent with the stellar activity for observed by Ribas *et al.* (2005) for the similarly aged G2V system EK Dra.

- The HD 61005 spectrum shows the observed emission can be separated into a hot, 7.5 MK stellar coronal component containing both continuum and line-emission contributions, coupled with a diffuse, extended, halo component dominated by line emission from hot plasma CXE.

- Taken all together, these results support a model of HD 61005 as a highly active, fast rotating young star with a strong stellar wind/spherical astropause.

- An intense, stellar-wind-ablating, via sputtering, HD 61005's ecliptic-plane, large cm-sized grains can create a stream of outflowing fine micron-sized particles which can be charged and picked up by the same wind. Once delivered outside the astropause by the stellar wind, these fine particles can be blown back by the instreaming ISM, creating the system's fine dust Wings. The same instreaming ISM can create a huge bubble of extended SWCX emission in HD 61005's astropause.

- Our Sun was in a similar state at times in its young life, when it was $\sim 10^8$ yrs old (Guinan & Engle 2007, Tu *et al.* 2015, Johnstone *et al.* 2021), and transiting through a dense part of the ISM such as a Giant Molecular Cloud (Stern 2003, Opher & Loeb 2024). Thus, models of our heliosphere that are informed by spacecraft measurements (e.g. see overviews in Dialynas *et al.* 2022, Galli *et al.* 2022, Kleiman *et al.* 2022) and drive the requirements for the future exploration of the heliosphere (e.g. Brandt *et al.* 2023), can be used to study this system as well, and our new Chandra measurements can be used to test and calibrate these models.

## 9. Acknowledgements.

This paper relies on data obtained by the Chandra X-ray Observatory (operated by the Smithsonian Astrophysical Observatory for and on behalf of the National Aeronautics Space Administration under contract NAS8-03060), and contained in dataset [DOI: 10.25574/cdc.352] {https://doi.org/10.25574/cdc.352}. We are indebted to the Chandra project for conducting further required testing, after the initial observations were obtained in 2021, of the energy-dependent quality of the Chandra ACIS-S PSF in 2023 and observing the system with HRC-I in 2024. C.M. Lisse gratefully acknowledges support for this work provided by the National Aeronautics and Space Administration through Chandra Award Number GO0-21018X issued by the Chandra X-ray Observatory Center (CXC). The authors would also like to thank T. Ayres, R. Cumbee, B. Draine, E. Gaidos, E. Mamajek, M. Opher, A. Poppe, and H. Throop for many useful discussions concerning X-ray emission from young dusty disk hosting sources. S. Deen aided this work greatly by performing an in-depth search for the latest information about HD 61005's edge-on inclination to the LOS and its non-binarity. D. Koutroumpa's contribution was supported by CNES, through its Sun-Heliosphere-Magnetosphere program. Kristina Kislyakova was supported by the European Union (ERC, EASE, 101123041. K. Dialynas was supported at JHU/APL by NASA under contracts NAS5 97271, NNX07AJ69G, and NNN06AA01C and by a subcontract at the Center for Space Research and Technology.

# 10. References.

## Appendix A.

### Chandra ACIS-S Imaging Instability (or why only E > 450 ev photon events were employed for this study).

The common wisdom in the Chandra observer community is that the observatory's HRMA telescope X-ray optics and X-ray image quality have been very stable since launch. However, this understanding has been achieved using monitoring of bright stellar point sources like AR Lac using the HRC camera (see the "Chandra Proposer's Observatory Guide", https://cxc.harvard.edu/proposer/POG/), not ACIS-S, and we are making a very strong claim about resolving HD 61005's Astrosphere that requires strong proof. ACIS-s's low-energy quantum efficiency/performance is known to have degraded substantially over time since launch (see "ACIS QE Contamination", https://cxc.cfa.harvard.edu/ciao/why/acisqecontamN0010.html). It is thought that increasing levels of an organic material have deposited onto the surface of the ACIS chips, producing increasing surface contamination over time that has substantially decreased the effective area performance of the instrument via increased X-ray absorption (for example, in the last two decades the effective collecting area at 0.3 keV has decreased by more than a factor of 20). This surface contamination could also have affected the ACIS-S imaging quality.

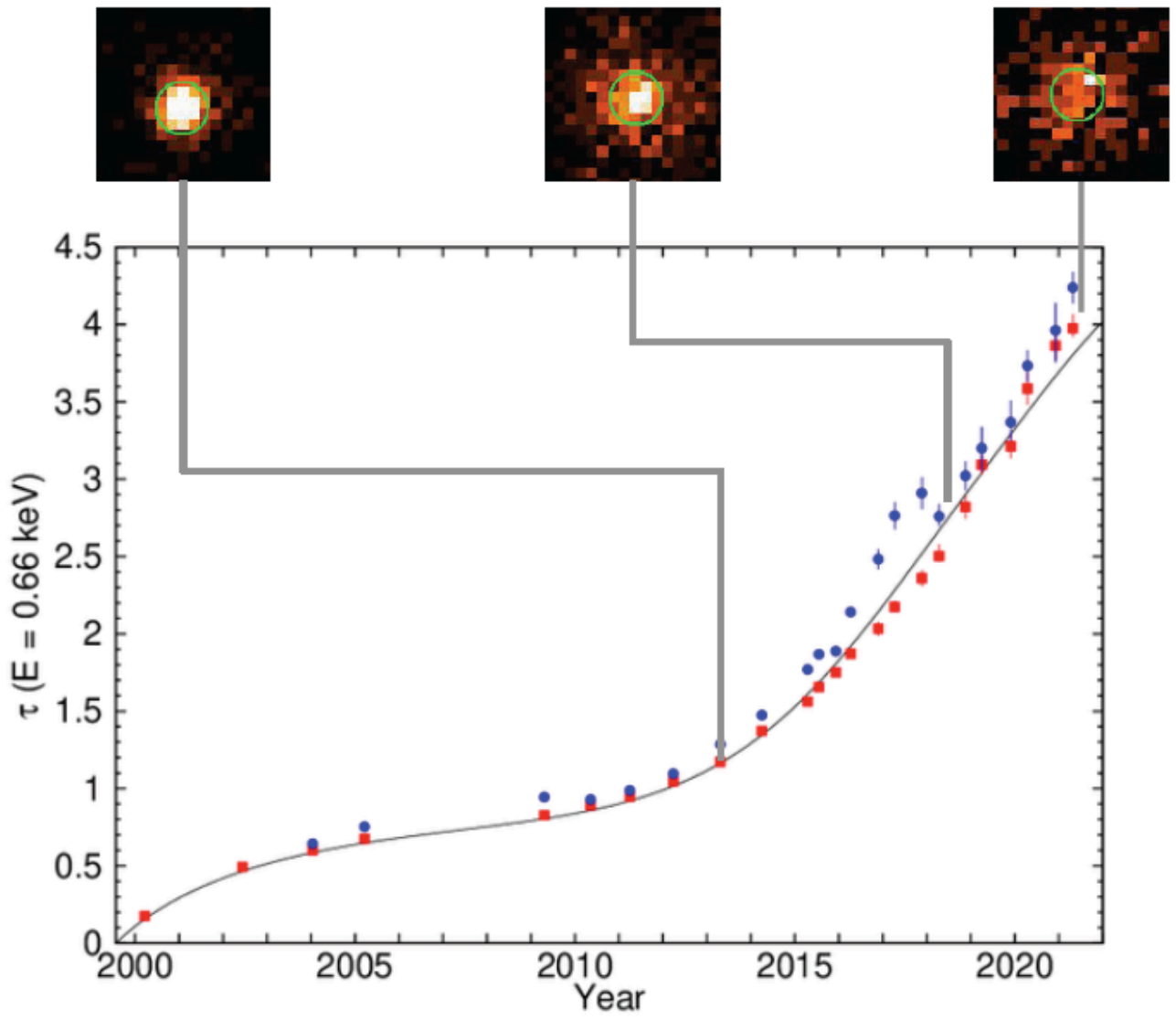

**Figure A1 - 0.1 to 0.35 keV image quality versus time for nearby, isolated (i.e., accretion disk-less) neutron star (NS) RXJ 1856.5-3754.** Here the median PSF quality at different times is denoted by the graphical figures at top, while the measured ACIS contamination optical depth is charted by the grey curve + red/blue points. Note that the NS is a very soft, point-like source, best able to provide a large # of event counts & good SNR for a 150 - 350 eV study.

In order to explicitly verify the validity of the Tau Ceti (observed in 2001) – Beta Hyi (observed in 2011) – HD 61005 (observed in 2021) G-star ACIS-S imaging comparison presented in this work, the image quality and amplitude of a known, bright, low-energy X-ray point source, the isolated/lone, nearby neutron star RX J18565_3457 (hereafter RXJ1856), was checked. Well studied by Chandra and XMM (e.g. Burwitz *et al.* 2001, 2003), this object evinces a very soft, featureless X-ray spectrum. Trending of the ACIS-S + LETG imaging grating spectroscopy performed every 2 years from 2011 to 2021 shows a marked degradation of the total count rate and expansion of the imaging PSF in the 150-300 eV energy range passband (Figure A1). This led to the harsh conclusion that differences in the 150 - 350 eV radial extension of HD 61005 as seen in 2021 versus that of Tau Ceti in 2001 cannot be utilized to conclude that the X-ray halo is much softer than the core.

RXJ1856 is a very soft X-ray object, making it difficult at first glance to determine the behavior of the ACIS-S PSF at higher energies. However, new careful analysis, coupled with a series of deep RXJ1856 calibration exposures taken by the Chandra project in 2023, has shown that the PSF is safely stable over time above 450 eV (see “Chandra Empirical PSF” discussion at https://cxc.harvard.edu/cal/Hrc/PSF/empPSF.html) and references therein. The working hypothesis is that ACIS image degradation is caused by increased x-ray scattering at energies where X-ray absorption and scattering is high – i.e. below the carbon and nitrogen K-alpha edges of atoms in the contaminant. Thus, the current PSF is most diffuse and extended at the lowest energies where the contaminant is most strongly absorbing, asymptoting to normal HRMA+ACIS-S behavior at the highest energies where the contaminants are X-ray transparent. This means that higher-energy photons, like those in HD 61005’s strong ~900 eV Ne IX halo charge exchange line, are not affected like 150-300 eV photons.

Concerning the use of isolated, low-ISM-interaction, stellar-calibration point sources for ACIS-S trending, Chandra has observed very few main sequence or giant branch stellar sources in general (only 80 stellar programs have been performed over 25 years of operation!). Even fewer trending studies of stellar point sources have been conducted over the two decades of its operation, especially of optically faint, but X-ray bright, targets that allow proper elucidation of its imaging quality trending at the lowest energies. The only consistently ACIS-S monitored objects we could find were Vega and Betelgeuse, and these stars are so optically bright that their spectra suffer strongly from optical loading effects. We were thus forced to use relative visit-by-visit comparison of the zeroth order ACIS+LETG grating imagery of RXJ1856 as a monitor because it is optically very faint but low-energy-X-ray bright, and was observed regularly over the course of the mission.

On the other hand, a measure of the PSF quality can be obtained from the 2021 HD 61005 Chandra ACIS data itself. Plotting the total, halo, core, and background event spectra on the same plot (Fig. A2), the background-corrected halo counts are unusually high, about the same number as the background-corrected core counts from 200 eV to 400 eV, and there is a marked spike in count rate at ~290 eV due to the Carbon K-edge. By contrast, starting at ~450 eV, we see that the halo counts steady down to ~20% of the core counts. This is the behavior we would expect for an ACIS-S PSF that is very diffuse near the carbon edge K-shell resonance at 285 eV, but is compact and reasonable above 450 eV.

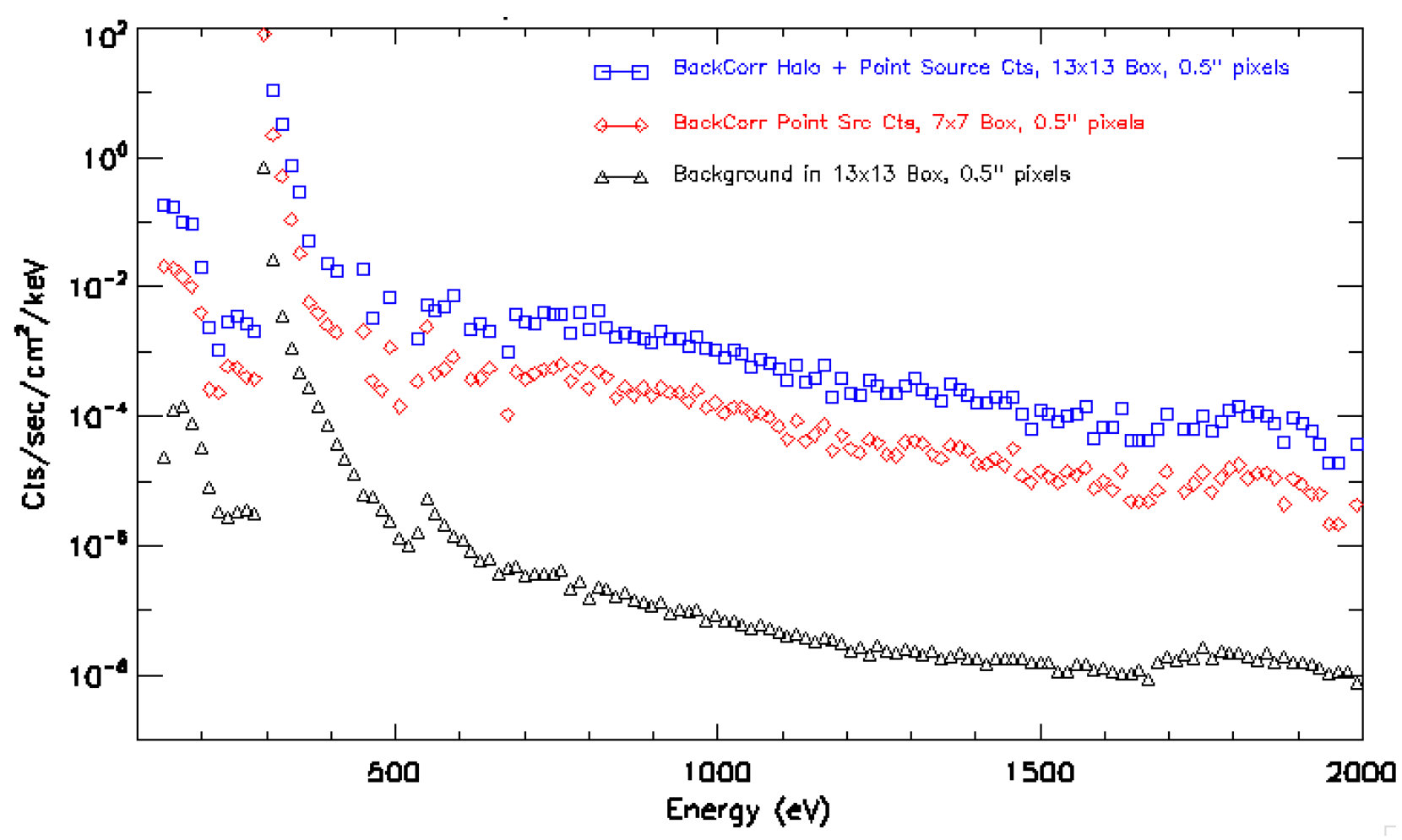

**Figure A2 - Background and source effective area corrected counts for the HD 61005 ACIS-S measurements taken in 2021.** Also included is a measure of the background spectrum. Note the irregular behavior of the halo (**red**) vs core (**blue**) events below 450 eV, the highly non-linear behavior around the Carbon K-shell edge at 283 eV, and that the background event rates (**black**) are more than 2 orders of magnitude lower than the source count rates above 450 eV.

Our finding of X-ray emission extended beyond that of a Chandra point source is corroborated by newly obtained HRC-I measurements of HD 61005 (Wolk *et al.* Nov 2024, OBSID 28004). The HRC-I does not suffer from the contamination effects of the ACIS-S camera, and has been used to monitor AR Lac, the Chandra PSF standard, to demonstrate unchanging PSF stability since launch in 2001. These measurements show that HD 61005 is indeed "fat" (or extended, with an approximately ~11% skirt of halo-like emission) versus the AR Lac radial profile, similar to the 12-15% excess halo emission found from our 2021 ACIS measurements (Fig. A3).

**Figure A3 – 0.1 to 2.0 keV "white light" HRC-I imaging observations of HD61005 and AR Lac, the Chandra imaging calibration standard. The** plot shows the aperture photometry for HD 61005 (diamonds) versus AR Lac (squares). (The curves have been normalized to match at r = infinity, as for Fig. 4a of the main text. Error bars are smaller than the symbols.) Out to ~1.5" from the center of brightness, HD 61005 evinces an extension over the AR Lac point source. The total amount of additional flux contained in this skirt is ~10% of the total AR Lac point source flux**,** consistent with the halo/core ratio found for HD 61005 using ACIS for E > 0.45 keV events reported above.

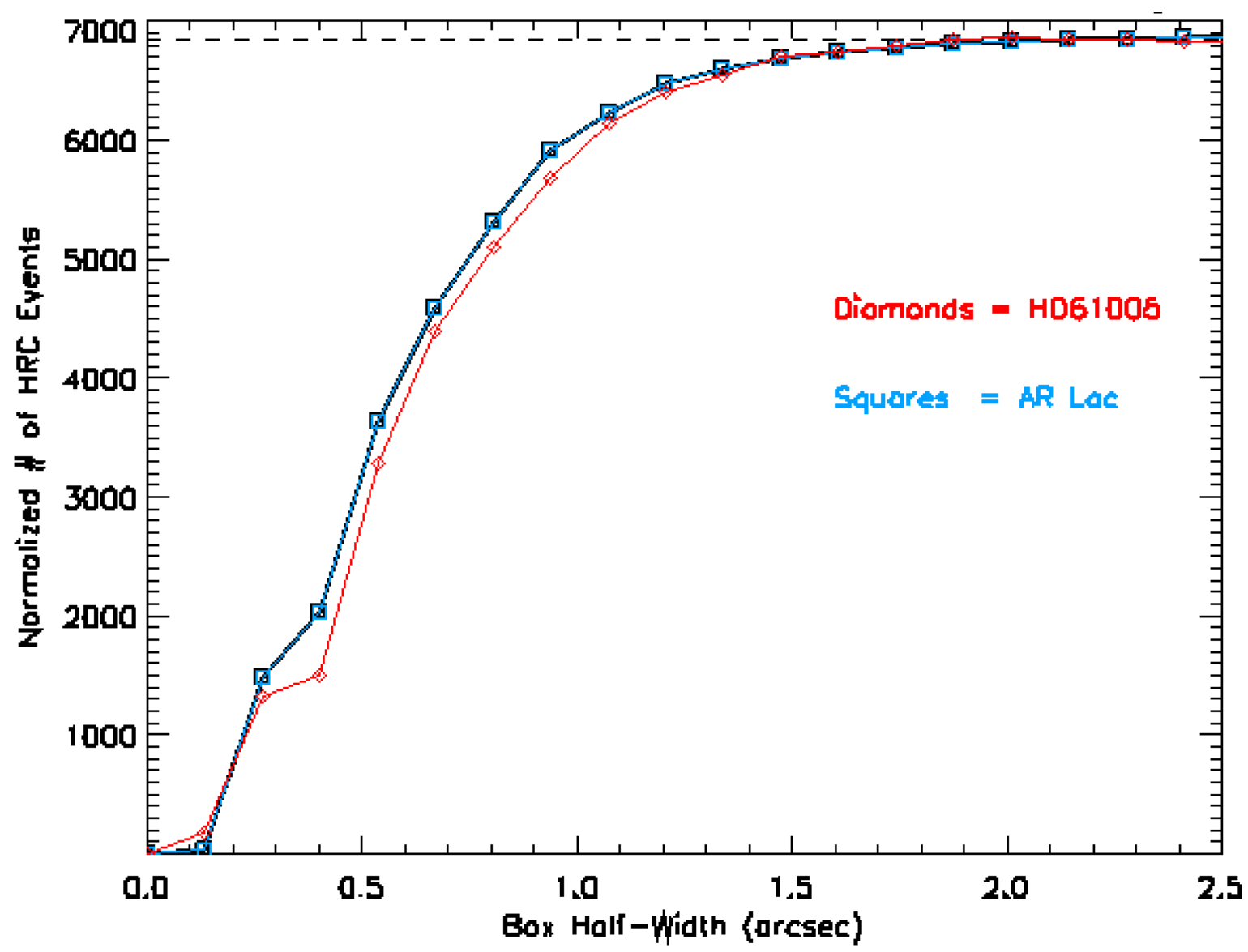

## Appendix B.

## Stellar Parameters Relevant to X-ray Emission for the G-Stars Studied in this Paper.*

| Object | Type | d (pc) | Age (Gyr) | log (Fe/H) | $P_{rot}$ (days) | $T_{corona}$ (MK) | $L_x$ (log erg $sec^{-1}$) | Chandra Observation Year(s) |
|---|---|---|---|---|---|---|---|---|
| | | | | | | | | |
| HD61005 | G9V | 36 | ~0.1 | ~0 | 5 | ~8 | 29.8 | 2021 |
| EK Dra | G0V | 34 | ~0.1 | ~0.05 | 4 | 2.27, 7.7 – 9.3 | 29.8 | 2001 |
| Sun | G2V | 0.0 | 4.56 | 0 | 27 | 1.22, 3.03 | ~27 | N/A |
| 18 Sco | G2V | 14 | ~4.5 | +0.04 | 22.5 | 1.5-2.0 | 26.4 | 2011 |
| Alpha Cen A | G2V | 1.3 | ~5.5 | +0.2 | 21 | ~1.2 | 27.1 | 2005 - 2024 |
| Tau Ceti | G8V | 3.7 | ~6 | -0.5 | 46 (?) | ~1.0?? | | 2001 |
| Beta Hyi | G2V | 7.5 | ~7 | -0.2 | ~28 | 2.15 | 27.5 | 2011 |

*After Ribas *et al.* 2005, 2015; Guinan and Engle 2007.